%% file: main_v3.tex
\documentclass[journal]{IEEEtran}

\usepackage[numbers]{natbib}
\usepackage[utf8]{inputenc} 
\usepackage[T1]{fontenc}    
\usepackage{hyperref}       
\usepackage{xurl}            
\usepackage{booktabs}       
\usepackage{subcaption}    
\usepackage{graphicx}      
\usepackage{amsfonts}       
\usepackage{nicefrac}       
\usepackage{algorithmicx}
\usepackage{algcompatible}
\usepackage{microtype}      
\usepackage{xcolor}         
\usepackage{amsmath}
\usepackage{amssymb}
\usepackage{algorithm}
\usepackage{algpseudocode}
\usepackage[english]{babel}
\usepackage{amsthm}
\usepackage{tabularx,booktabs,siunitx}
\usepackage{adjustbox}  
\usepackage[english]{babel}
\usepackage{amsthm}

\newtheorem*{remark}{Remark}
\input{macros.tex}
\input{acronyms.tex}

\begin{document}

\title{RSS map-assisted MIMO channel estimation in the upper mid-band under pilot constraints}

\author{Alireza~Javid,~\IEEEmembership{Student Member,~IEEE,}
        and~Nuria~Gonz\'alez-Prelcic,~\IEEEmembership{Fellow,~IEEE}
        \thanks{An earlier version of this paper was accepted to the 2026 IEEE International Conference on Acoustics, Speech, and Signal Processing \cite{Javid2025arXiv}. 
A. Javid and  N. Gonz\'{a}lez-Prelcic are with the Department of Electrical and Computer Engineering, University of California San Diego, USA (e-mail: \{sajavid,ngprelcic\}@ucsd.edu).}
}

\maketitle

\begin{abstract}
Accurate wireless channel estimation is critical for next-generation wireless systems, enabling precise precoding for effective user separation, reduced interference across cells, and high-resolution sensing, among other benefits. 
Traditional model-based channel estimation methods suffer, however, from performance degradation in complex environments with a limited number of pilots, while purely data-driven approaches lack physical interpretability, require extensive data collection, and are usually site-specific. This paper presents a novel \ac{PINN} framework that synergistically combines model-based channel estimation with a deep network to exploit prior information about environmental propagation characteristics and achieve superior performance under pilot-constrained scenarios. The proposed approach employs an enhanced U-Net architecture with transformer modules and cross-attention mechanisms to fuse initial channel estimates with \ac{RSS} maps to provide refined channel estimates. Comprehensive evaluation using realistic ray-tracing data from urban environments demonstrates significant performance improvements, achieving over 5 dB gain in \ac{NMSE} compared to state-of-the-art methods, with particularly strong performance in pilot-limited scenarios  and robustness across different frequencies and environments with only minimal fine-tuning. We further extend the decoder for multi-step temporal prediction, enabling accurate forecasting of several future channel snapshots from a single estimate—useful for proactive beamforming and scheduling in mobile scenarios. The proposed framework maintains practical computational complexity, making it viable for massive \ac{MIMO} systems in upper-mid band frequencies. Unlike black-box neural approaches, the physics-informed design provides a more interpretable channel estimation method.
\end{abstract}


\IEEEpeerreviewmaketitle

\section{Introduction}
Accurate \ac{CSI} underpins nearly every physical-layer function in multi-antenna systems. It drives precoding and combining, user scheduling, link adaptation (modulation and coding selection), and equalization \cite{ozdemir2007channel}, and is increasingly reused for localization and sensing \cite{palacios2023separable}. The need for high-fidelity \ac{CSI} is amplified with large arrays, multiuser spatial multiplexing, and wideband operation, where narrow beams, inter-user coupling, and frequency selectivity make the system highly sensitive to estimation errors. Inaccurate \ac{CSI} manifests as beam misalignment, residual multiuser interference, suboptimal rate selection, and ill-conditioned equalizers, ultimately reducing SINR, spectral efficiency, reliability, and sensing resolution. These challenges are further exacerbated under mobility and hardware constraints, where pilot overhead, feedback latency, and computational limits restrict how often and how precisely \ac{CSI} can be acquired. Moreover, the emerging upper mid-band (FR3) spectrum—typically spanning roughly 7–24 GHz—offers a favorable balance between bandwidth availability and propagation resilience compared to conventional sub-6 GHz and millimeter-wave bands \cite{bazzi2502upper}. However, operating in FR3 brings new challenges: larger path loss, increased material penetration losses, and more pronounced frequency selectivity demand higher estimation fidelity, especially across wideband channels \cite{abbasi2025ultra}. This paper will focus on channel estimation for this band as the future of wireless systems.
\subsection{Related works}
Traditional model-based channel estimation techniques for large-array MIMO systems often rely on pilots. However, as the number of antennas increases, they require excessive training and incur high computational complexity, which leads to performance degradation under overhead constraints or in complex propagation environments \cite{palacios2022multidimensional, palacios2023separable, venugopal2017channel, rodriguez2018frequency}. This problem is expected to intensify with the large-array configurations envisioned for FR3 systems, where wider bandwidths and higher antenna counts further amplify pilot overhead and computational demands \cite{bazzi2502upper}.
Prior work has also proposed black-box, fully data-driven approaches \cite{sattari2024full, helmy2025low, hu2020deep, zhou2025generative} that often suffer from limited physical interpretability, demand large-scale data collection, or tend to generalize poorly beyond the specific site or frequency band where they were trained.

Another line of work combines model-based priors with data-driven learning \cite{jin2025near}. However, these combined approaches often utilize the black-box structure of neural networks along with simplistic channel models and unrealistic statistical data for training. Consequently, performance often degrades under domain (site) shift, and substantial site-specific data are required for training when the deployment location changes. In addition, the exploitation of diffusion models in the recent channel estimation literature \cite{jin2025near, zhou2025generative}  will also introduce a high overhead in the backward process for generating the channels, and cannot be used in real-time communication systems. \cite{qiu2025ai} also assumes narrowband single-antenna UEs and relies on location-specific denoisers and iterative MAP solvers. In contrast, we address wideband multi-tap MIMO with physics-informed RSS maps, enabling single-shot inference and generalization without \ac{CKM} databases. 

\ac{PINN} have emerged as a powerful paradigm that integrates domain knowledge from physical laws into neural network architectures, enabling more interpretable and data-efficient learning \cite{raissi2019physics, karniadakis2021physics}. By incorporating governing equations, boundary conditions, and conservation laws directly into the loss function or network structure, PINNs have demonstrated remarkable success across diverse scientific computing applications, including fluid dynamics, structural mechanics, and electromagnetic wave propagation \cite{wang2021learning,jin2021nsfnets}. Unlike purely data-driven approaches, PINNs leverage prior physical knowledge to constrain the solution space, resulting in improved generalization with limited training data and enhanced robustness to out-of-distribution scenarios. In the context of wireless communications, recent works have begun exploring physics-informed approaches for channel modeling and parameter estimation \cite{wang2025physics, jiang2024physics}, with some successfully incorporating detailed electromagnetic field calculations and environmental propagation characteristics through ray tracing methods and integral equation formulations. However, these approaches often focus on specific aspects of channel modeling, such as regional channel impulse response estimation or path loss prediction, while the integration of physics-informed constraints with pilot-limited channel estimation for massive \ac{MIMO} systems remains relatively unexplored, particularly in scenarios requiring both high accuracy and efficiency under severe pilot constraints.

Building on our previous work \cite{Javid2025arXiv}, this paper further develops a framework that combines traditional model-based wireless channel estimation with a physics-informed neural network (PINN) approach. This hybrid strategy leverages environmental information to reduce overhead and enhance performance while avoiding the end-to-end black-box structure. Additionally, we evaluate our approach using ray tracing data, which closely resembles real-world data and effectively captures most characteristics of the environment \cite{skidmore2021auto}. Numerical results show an improvement of over 5 dB compared to the baselines, while having a practical latency for real-time applications. We also demonstrated the capability of this method on different frequency bands and environments. Beyond computational efficiency, this approach offers physical interpretability that distinguishes it from purely black-box neural methods. Additionally, for mobile scenarios, we extend the decoder into a parallel multi-step temporal head that estimates $L$ future channel snapshots from a single input, avoiding autoregressive error accumulation while preserving physical consistency with the RSS constraints. 

\subsection{Contributions}
This paper presents a novel \ac{PINN} framework for wireless channel estimation under pilot-constrained scenarios, with the following key contributions: 
\begin{itemize}
    \item 
     We propose a hybrid architecture that synergistically combines model-based channel estimation with deep learning by incorporating environmental propagation characteristics through \ac{RSS} maps as input. This approach leverages prior physical knowledge to constrain the solution space even when pilot signals are severely limited.
     \item 
     We design an enhanced U-Net architecture augmented with transformer modules and cross-attention mechanisms that effectively fuse initial channel estimates with environmental information, enabling the network to selectively incorporate physics-based environmental knowledge rather than treating channel estimation and propagation characteristics as independent modalities. To the best of our knowledge, this is the first work to integrate RSS maps into a physics-informed neural network for pilot-limited MIMO channel estimation.
     \item 
     We construct several realistic ray-tracing datasets spanning multiple carrier frequencies and deployment sites to evaluate the proposed approach and assess its generalization across distribution shifts. This dataset is also provided for future contributions. 
     \item 
     We demonstrate performance improvements through comprehensive evaluation on realistic ray-tracing datasets from urban environments, achieving over 5 dB gain in normalized mean squared error compared to state-of-the-art methods, with particularly strong performance in pilot-limited scenarios where our approach achieves approximately -13 dB NMSE with only four pilots at 0 dB signal-to-noise ratio. The codes and experiments are also available for future research. 
      \item 
We extend the framework to enable multi-step temporal channel estimation, allowing the network to forecast multiple future channel snapshots from a single current coarse estimate. This capability enables proactive resource allocation and beamforming adaptation in dynamic wireless environments while maintaining computational efficiency.
      
\end{itemize}
\subsection{Organization}
The rest of the paper is organized as follows. Section~\ref{model} presents the physical design, system model, and problem formulation. Section~\ref{nndisc} describes the proposed physics-informed neural network, including initial estimation, physical calculations, and network design. Section~\ref{data} introduces the dataset, while Section~\ref{exps} provides experimental setup, complexity analysis, and performance evaluation, including generalization studies. Section~\ref{con} concludes the paper.
\subsection{Notation}
$\bX^{\top}$ and $\bX^H$ are the transpose and Hermitian of matrix $\bX . [\bX, \bY]$ and $[\bX ; \bY]$ are the horizontal and vertical concatenation of $\bX$ and $\bY$. $\bX \otimes \bY$ is the Kronecker product of $\bX$ and $\bY$.  We use $ \bbC^{D \times 1}$ and $\bbR^{D \times 1}$ to represent the $D$-dimensional space of complex and real-valued vectors, respectively. We also use $\|\cdot\|$ to denote the $L^2$-norm, which is an Euclidean norm.

\section{Models and Problem Formulation} \label{model}
\subsection{Physical model} \label{sec:phy}
 Maxwell's equations form the theoretical foundation for wireless signal propagation, governing the behavior of electromagnetic waves as they interact with the environment \cite{ida2015engineering}. These fundamental equations describe the relationships between electric and magnetic fields, permittivity, permeability, and current densities in spatially varying media. 

Under the time-harmonic assumption and appropriate simplifications, Maxwell's equations can be reduced to an inhomogeneous wave equation for the electric field, where the position-dependent wavenumber accounts for material property variations throughout the propagation environment \cite{ida2015engineering}. The solution to this wave equation can be obtained using Green's function methods, which provide the fundamental solution for electromagnetic wave propagation in inhomogeneous media.

The Green's function approach allows us to express the electric field as a superposition of contributions from source currents and medium inhomogeneities. This formulation requires detailed knowledge of material properties such as wall thickness, relative permittivity and permeability, and electrical conductivity, which determine how electromagnetic waves are propagated at material boundaries.

After computing the electric field distribution using numerical methods, the total received power at a given location can be determined from the electric field. Specifically, the received power is proportional to the squared magnitude of the electric field integrated over the receiving antenna's effective area. Mathematically, the total received power at the receiver, accounting for all \ac{MPC}, can be expressed as \cite{fu2023fast}
\begin{equation}
	 P_R^{\text{EM}} = \frac{\lambda^2}{8 \pi \eta_0} \left| \sum_{\ell=1}^{P} E_{\ell}(\mathbf{r}) \right|^2,
	\label{eq:received_power}
\end{equation}
where $P$ denotes the number of propagation paths, $\lambda$ is the wavelength, $\eta_0$ is the intrinsic impedance of free space, where $\mathbf{r} \in \mathbb{R}^3$ denotes the receiver position in global coordinates, and $E_{\ell}$ represents the complex amplitude of the electric field associated with the $\ell$-th path at the receiver location 
. By defining $E(\mathbf{r}) = \sum_{\ell=1}^{P} E_{\ell}(\mathbf{r})$, which captures the coherent superposition of all path contributions, including their respective amplitudes and phases,  we can express $P_R^{\text{EM}} = \frac{\lambda^2}{8 \pi \eta_0} \left|E(\mathbf{r})\right|$. This relationship  connects numerical computations of electric fields with received signal strength measurements in wireless communication systems.

\subsection{Communication system model} \label{sec:comm}
 We consider the downlink of a communication system operating in upper-mid bands (7-24 GHz), in an urban environment. The \ac{BS}, equipped with a \ac{URA} of size $N_t = N_t^x \times N_t^y$, is at the top of a building, and the receiver is located in a vehicular user and equipped with a \ac{URA}  of $N_r = N_r^x \times N_r^y$ elements.  We consider a time-domain frequency-selective channel due to multipath propagation. Hereby, the frequency-selective channel consisting of ${P}$ paths after sampling from a continuous-time channel can be defined as
\begin{equation}
\label{channel}
\mathbf{H}_d=\sum_{\ell=1}^{P}\alpha_{\ell} f_{\mathrm{p}}\left(d T_{\mathrm{s}}-\left(t_{\ell}-t_{\mathrm{off}}\right)\right)\mathbf{a}_{\mathrm{r}}\left(\theta_{\ell}^{\mathrm{az}}, \theta_{\ell}^{\mathrm{el}}\right) \mathbf{a}_{\mathrm{t}}\left(\phi_{\ell}^{\mathrm{az}}, \phi_{\ell}^{\mathrm{el}}\right),
\end{equation}
where $d$ is the channel tap index, $T_\textrm{s}$ is the sampling interval, $t_{\mathrm{off}}$ is the clock offset between the transmitter and receiver, $f_{\mathrm{p}}(.)$ is the filtering function that factors in filtering effects in the system, $\alpha_{\ell}$ and $t_{\ell}$ are the complex gain and the \ac{ToA} of the $l$-th path, $\mathbf{a}_{\mathrm{r}}\left(\theta_{\ell}^{\mathrm{az}}, \theta_{\ell}^{\mathrm{el}}\right)$ represents the receiver array response evaluated at the azimuth and elevation \ac{AoA}, denoted as $\theta_{\ell}^{\mathrm{az}}$ and $ \theta_{\ell}^{\mathrm{el}}$ respectively, and $ \mathbf{a}_{\mathrm{t}}\left(\phi_{\ell}^{\mathrm{az}}, \phi_{\ell}^{\mathrm{el}}\right)$ is the transmitter array response evaluated at the azimuth and elevation \ac{AoD}, denoted as $\phi_{\ell}^{\mathrm{az}}$ and $ \phi_{\ell}^{\mathrm{el}}$ respectively. The transmit and receive array responses can be formulated as
\begin{subequations}
\begin{align}
\mathbf{a}_{\mathrm{r}}\left(\theta^{\mathrm{az}}, \theta^{\mathrm{el}}\right)=\mathbf{a}\left(\theta^{\prime \prime}, \theta^{\perp}\right)=\mathbf{a}\left(\theta^{\prime \prime}\right) \otimes \mathbf{a}\left(\theta^{\perp}\right), \\
\mathbf{a}_{\mathrm{t}}\left(\phi^{\mathrm{az}}, \phi^{\mathrm{el}}\right)=\mathbf{a}\left(\phi^{\prime \prime}, \phi^{\perp}\right)=\mathbf{a}\left(\phi^{\prime \prime}\right) \otimes \mathbf{a}\left(\phi^{\perp}\right),
\end{align}
\end{subequations}
where $\theta^{\prime \prime}=\cos \left(\theta^{\mathrm{el}}\right) \sin \left(\theta^{\mathrm{az}}\right), \theta^{\perp}=\sin \left(\theta^{\mathrm{el}}\right), \phi^{\prime \prime}=$ $\cos \left(\phi^{\mathrm{el}}\right) \sin \left(\phi^{\mathrm{az}}\right), \phi^{\perp}=\sin \left(\phi^{\mathrm{el}}\right)$, and $\mathbf{a}(\cdot)$ is the steering vector where $[\mathbf{a}(\vartheta)]_n=e^{-j \pi(n-1) \vartheta}$ assuming a half-wavelength element spacing.

\subsection{Physical modeling and communication channel} The electromagnetic model  described in Section~\ref{sec:phy} provides a physically grounded approach for modeling wireless propagation by explicitly computing the electric field at any location, accounting for the detailed properties of the environment. In contrast, modern communication system models represent the channel by a finite set of propagation paths, each parameterized by delay, angle, and complex gain, as encapsulated in the \ac{MIMO} channel matrix $\{\mathbf{H}_d\}_{d=1}^D$ defined in Section~\ref{sec:comm}, each corresponding to a delay tap $d$, with $D$ the total number of resolvable taps. 
The physically computed field superposition at each receiver location thus defines the per-tap response in the wideband \ac{MIMO} channel. The received signal strength can be calculated per tap, or integrated across all taps, using the \ac{MIMO} channel matrices, as
\begin{equation}
P_R^{\text{Chan}} = P_T \sum_{d=1}^{D} \mathbb{E}\left[\|\mathbf{H}_d\|_F^2\right]
\end{equation}
or, equivalently, from the summed field strengths at each tap. Here, $P_T$ is the transmission power. 


\subsection{Problem Formulation}

Given the received signal model, the objective is to estimate the frequency-selective MIMO channel $\mathbf{H} = \{\mathbf{H}_d\}_{d=0}^{D-1}$ from $N_p$ pilot under noisy measurement.

The limited number of pilots introduces an underdetermined estimation problem, where the number of channel unknowns significantly exceeds the available measurements. Classical channel estimation methods perform poorly in this regime.

We address this by learning a refinement mapping that combines an initial coarse estimate $\tilde{\mathbf{H}}$ with prior physical information about the propagation environment encoded through RSS maps, producing an improved estimate $\hat{\mathbf{H}}$ that minimizes NMSE while maintaining consistency with electromagnetic propagation principles.

\section{Physical Informed Neural Network} \label{nndisc}
\begin{figure}[t]
	\centering
	\includegraphics[width=\columnwidth]{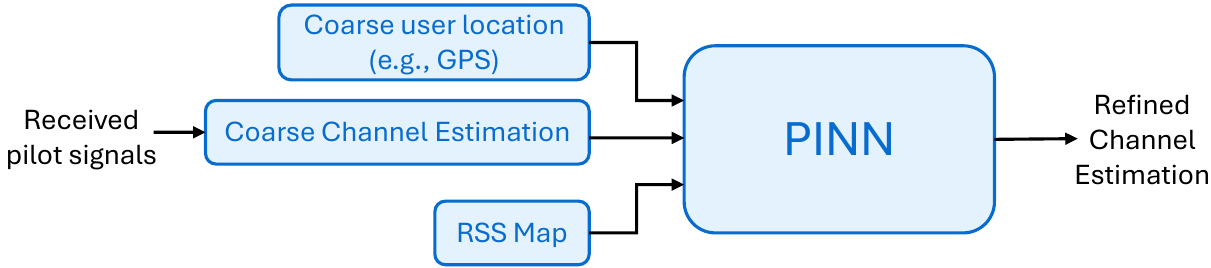}
	\caption{Block diagram of the proposed PINN-based channel estimation approach.}
	\label{fig:block_diagram}
\end{figure}

The proposed approach combines model-based channel estimation with learned refinement using environmental prior information. Figure~\ref{fig:block_diagram} illustrates the overall processing flow. 

The system takes received pilot signals as input and produces three parallel inputs for the neural network: (i) a coarse channel estimate obtained from classical methods described in Section~\ref{sec:initial_est}, (ii) coarse user location information, and (iii) an RSS map computed from electromagnetic propagation analysis as detailed in Section~\ref{sec:phy}. The PINN then fuses these inputs to produce a refined channel estimate, leveraging the physical consistency between channel characteristics and environmental propagation patterns. The network architecture is presented in Section~\ref{sec:nn_design}.

\subsection{Initial channel estimation} The first step consists of a  coarse channel estimation. To simplify our approach and avoid unnecessary complications, we utilize basic variations of \ac{LS} estimation. However, our overall \ac{PINN} solution can include other initial channel estimation methods. 

The frequency-selective \ac{MIMO} channel is modeled as a 3-dimensional complex tensor $\mathbf{H} \in \mathbb{C}^{D \times N_{\text{r}} \times N_{\text{t}}}$. The received signal can be written as
\begin{equation}
\mathbf{r}[n]= \sum_{d=0}^{D-1} \mathbf{H}_d \mathbf{s}[n-d]+\mathbf{v}[n],
\end{equation}
where $\mathbf{s}[n] \in \mathbb{C}^{N_{\text{t}}}$ is the transmitted pilot vector and $\mathbf{v}[n] \sim \mathcal{CN}(\mathbf{0}, \sigma_n^2 \mathbf{I})$ is the additive white Gaussian noise.

\subsubsection{LS with Linear Interpolation} \label{sec:initial_est}
This method places pilots at a subset of transmit antenna indices. The pilot set is defined as
\begin{equation}
	\mathcal{P} = \{p_i : p_i = i \cdot \Delta_p, \quad i = 0, 1, \ldots, N_p - 1\},
\end{equation}
where $\Delta_p = \max(1, \lfloor N_t / N_p \rfloor)$ is the antenna spacing between pilots. The received signal at pilot antenna $p_i$ for tap $d$ and receive antenna $r$ is
\begin{equation}
	y_{d,r}[p_i] = h_{d,r}[p_i] + n[p_i],
\end{equation}
where $n \sim \mathcal{CN}(0, \sigma_n^2)$. LS estimation yields $\hat{h}_{d,r}[p_i] = y_{d,r}[p_i]$. The full channel vector $\hat{\mathbf{h}}_{d,r} \in \mathbb{C}^{N_t}$ is obtained via linear interpolation of magnitude and unwrapped phase across all $N_t$ antenna positions.

\subsubsection{LS with DFT Denoising}
This method exploits channel sparsity in the beamspace domain for noise reduction. For each delay tap $d \in \{0, \ldots, D-1\}$ and receive antenna $r \in \{1, \ldots, N_r\}$, the channel vector $\mathbf{h}_{d,r} \in \mathbb{C}^{N_t}$ represents the channel coefficients across all $N_t$ transmit antennas. After obtaining \ac{LS} estimates at pilot antenna positions and zero-padding non-pilot positions, the \ac{DFT} transforms the spatial-domain channel to the beamspace domain:
\begin{equation}
	\mathbf{H}_{d,r}^{(\text{beam})} = \mathbf{F}^H \tilde{\mathbf{h}}_{d,r}
\end{equation}
where $\mathbf{F} \in \mathbb{C}^{N_t \times N_t}$ is the normalized \ac{DFT} matrix and $\tilde{\mathbf{h}}_{d,r} \in \mathbb{C}^{N_t}$ is the zero-padded pilot-based estimate (nonzero only at pilot antenna indices). Since massive MIMO channels exhibit angular sparsity, adaptive thresholding removes weak beamspace components:
\begin{equation}
	\hat{H}_{d,r}^{(\text{beam})}[k] = \begin{cases}
		H_{d,r}^{(\text{beam})}[k] & \text{if } |H_{d,r}^{(\text{beam})}[k]| \geq \tau \\
		0 & \text{otherwise}
	\end{cases}
\end{equation}
where $k \in \{0, \ldots, N_t - 1\}$ indexes the beamspace bins and the threshold is
\begin{equation}
	\tau = 3\sqrt{\frac{\sigma_n^2 N_p}{2N_t}}.
\end{equation}
The denoised channel is reconstructed via inverse \ac{DFT}:
\begin{equation}
	\hat{\mathbf{h}}_{d,r} = \mathbf{F} \hat{\mathbf{H}}_{d,r}^{(\text{beam})}.
\end{equation}
This method effectively reduces noise by removing weak beamspace components while preserving dominant angular directions \cite{rao2014distributed}.
\subsubsection{LS OFDM-Based Estimation}
This method operates in the frequency domain using \ac{OFDM} subcarrier structure. Pilot subcarrier positions are defined with uniform spacing,
\begin{equation}
\mathcal{S} = \{s_i : s_i = i \cdot \Delta_s, \quad i = 0, 1, ..., N_s-1\}.
\end{equation}
The time-domain channel is transformed to the frequency domain with zero-padding as
\begin{equation}
\mathbf{H}^{\text{(freq)}}[k] = \text{FFT}(\mathbf{h}_{d,r}, N_{\text{subcarriers}})[k],
\end{equation}

After LS estimation at pilot subcarriers, we interpolate to data subcarriers using linear interpolation for magnitude and phase \cite{ozdemir2007channel} 
{\scriptsize
\begin{equation}
\hat{H}^{\text{(freq)}}[k] = \begin{cases}
\hat{\mathbf{H}}_{\text{pilots}}[i] & k \in \mathcal{S} \\
|\text{interp}(\mathcal{S}, |\hat{\mathbf{H}}_{\text{pilots}}|, k)| \cdot e^{j\text{interp}(\mathcal{S}, \text{unwrap}(\angle\hat{\mathbf{H}}_{\text{pilots}}), k)} & k \notin \mathcal{S}
\end{cases}.
\end{equation}
}
The time-domain channel is recovered using \ac{IFFT}, retaining only the first $D$ delay taps:
\begin{equation}
\hat{\mathbf{h}}_{d,r} = \text{IFFT}(\hat{\mathbf{H}}^{\text{(freq)}} , N_{\text{subcarriers}})[1:D].
\end{equation}
This method yields the best initial results compared to other methods, although its complexity is higher. The complexity of these methods is summarized in Table~\ref{tab:complexity-comparison}.
\begin{table}[t]
  \centering
  \caption{Computational complexity of initial channel estimation methods}
  \label{tab:complexity-comparison}
  \begin{tabular}{ll}
    \toprule
    \textbf{Method} & \textbf{Complexity} \\
    \midrule
    LS + linear interp. & $\mathcal{O}\!\left(DN_r(N_p+N_t)\right)$ \\
    LS + DFT denoising & $\mathcal{O}\!\left(DN_rN_t\log N_t\right)$ \\
    LS-OFDM (freq.-dom.) & $\mathcal{O}\!\left(DN_rN_t\log N_{\text{FFT}} + N_rN_tN_{\text{FFT}}\right)$ \\
    \bottomrule
  \end{tabular}
\end{table}
\subsection{Physical calculation of the received power} The physically informed network to be designed, exploits the \ac{RSS} map obtained using the numerical methods discussed earlier. A popular non-neural simulator for this computation is \textit{Wireless Insite} \cite{remcom2022wireless}, which can accurately describe the physical structure of the environment. Section~\ref{data} contains more information for this stage. We combine the behavior of the wireless channel with physical calculations through a physically informed neural network. We assume that the BS has a rough estimate of the user's position. This information helps locate users on the \ac{RSS} map, allowing us to extract the approximate power levels in those areas for use in the network. In practical wireless deployments, \ac{RSS} maps can be evaluated using digital twin technology, which creates real-time virtual representations of the electromagnetic propagation environment. Commercial digital twin platforms like VIAVI's Network Digital Twin and Ericsson's NDT solutions already provide AI-driven radio propagation modeling that can generate accurate RSS maps synchronized with physical network conditions \cite{ericsson2023digitaltwin}, while advanced ray tracing tools such as NVIDIA's Instant Radio Maps can compute high-resolution radio maps using real-world 3D environmental data \cite{10684249}. These digital twins leverage high-definition 3D maps and ray-based propagation simulation to provide real-time digital representations of electromagnetic environments.

\subsection{RSS Feature Extraction}

The RSS map provides spatially distributed information about electromagnetic propagation characteristics in the environment. To leverage this information effectively for channel estimation, we extract a localized RSS feature representation centered on the user's estimated position.

Given the user's position estimate $(\hat{x}_u, \hat{y}_u)$, we crop a $W \times W$ meter region from the precomputed RSS map. This cropping operation accounts for GPS positioning uncertainty by capturing the RSS distribution in the user's vicinity rather than relying on a single-point measurement. The cropped region is denoted as $\mathbf{R}_u \in \mathbb{R}^{H_c \times W_c}$, where $H_c$ and $W_c$ are the pixel dimensions corresponding to the physical crop size.

The RSS encoder $f_\text{RSS}(\cdot)$ processes the cropped map through a series of convolutional blocks to extract a compact feature representation:
\begin{equation}
	\mathbf{z}_\text{RSS} = f_\text{RSS}(\mathbf{R}_u) \in \mathbb{R}^{D_z},
\end{equation}
where $D_z$ is the latent dimension. The encoder architecture consists of four convolutional blocks, ReLU activation, and max-pooling, progressively increasing the channel depth while reducing spatial dimensions. A final adaptive average pooling layer produces the fixed-length feature vector regardless of input resolution variations.

This hierarchical feature extraction captures multi-scale spatial patterns in the RSS distribution, from local signal variations to broader propagation characteristics such as shadowing regions and dominant reflection paths. The extracted features $\mathbf{z}_\text{RSS}$ are subsequently integrated with channel features through the cross-attention mechanism described in Section~\ref{sec:nn_design}, enabling the network to selectively incorporate environment-specific propagation knowledge into the channel estimation process.

\subsection{Neural network design}
\label{sec:nn_design}
\begin{figure*}[!h]
    \centering
    \includegraphics[width=.9\linewidth]{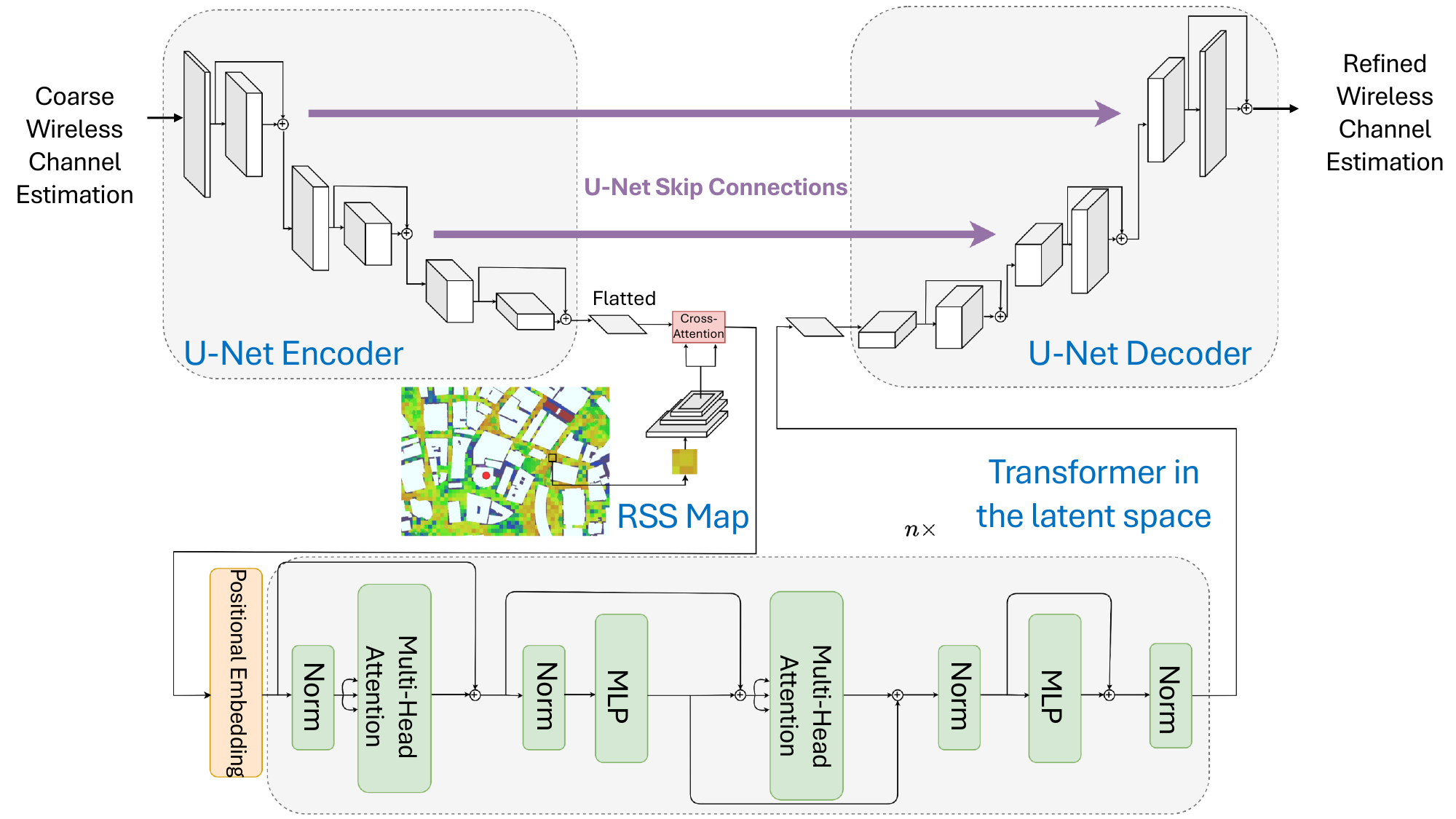}
    
    \caption{\ac{PINN} structure for channel estimation.}
    \label{fig:nn}
\end{figure*}

 We employ a physics-informed U-Net architecture with transformer enhancement as illustrated in Figure \ref{fig:nn}. The network consists of a three-layer encoder-decoder Resnet-based structure with symmetric skip connections, specifically designed for the channel estimation task with spatial dimensions $(N_r, N_t)$. The encoder progressively downsamples the $2\times D$-channel input through convolutional blocks, reducing the spatial resolution while increasing the feature depth.
In the latent space, after flattening the extracted features, we integrate the physics-informed \ac{RSS} map information through a dedicated \ac{RSS} encoder that processes the cropped environmental map around the user location. Cross-modal fusion combines the flattened channel features with the \ac{RSS} features via a cross-attention that learns the correlation between electromagnetic propagation characteristics and channel structure. The cross-attention mechanism integrates \ac{RSS} environmental information with channel features. This process started with a projection to a common hidden dimension as
\begin{equation}
\mathbf{X}_i = \mathbf{W}_i \mathbf{F}_i + \mathbf{b}_i\:\:; \:\: i \in \{\text{\small{RSS, Channel}}\}, 
\end{equation}
in which $\mathbf{F}_i \in \mathbb{R}^{\text{Batch} \times D_i}$ is the extracted features from \ac{RSS} and channel, $\mathbf{X}_i \in \mathbb{R}^{\text{Batch} \times D_z}$ is the projected features, $\mathbf{W}_i  \in \mathbb{R}^{D_i \times D_z}$ is the projection matrices, and $\mathbf{b}_i\in \mathbb{R}^{ D_z}$ is the bias vector. Here, $D_z$ represents the dimension of the latent space, and $D_i$ is the dimension of \ac{RSS} or channel features. The cross-attention mechanism employs multi-head attention. The process is formulated as $\text{softmax}\left(\frac{\mathbf{Q}\mathbf{K}^T}{\sqrt{D_z}}\right)\mathbf{V}$ where $\mathbf{Q} = \mathbf{X}_{\text{Channel}}$, $\mathbf{K} = \mathbf{X}_{RSS}$, and $\mathbf{V} = \mathbf{X}_{RSS}$. Setting $\mathbf{Q}$ as channel features and $\mathbf{K}$, $\mathbf{V}$ as \ac{RSS} features enables the channel estimation to query what environmental information is most relevant for refinement, where the attention weights determine which \ac{RSS} spatial patterns are most informative for each channel element. This cross-modal attention allows the network to selectively incorporate physics-based environmental knowledge into the channel estimation process rather than treating them as independent modalities.
The fused features are then processed by a transformer module consisting of several self-attention blocks, where the spatial positions are treated as a sequence. This transformer component enables the network to model long-range dependencies between different channel taps and capture the physics-informed relationships between \ac{RSS} patterns and channel characteristics. The decoder mirrors the encoder structure, utilizing transposed convolutions to progressively upsample the transformed features back to the original channel dimensions. Skip connections from corresponding encoder layers preserve fine-grained details and facilitate gradient flow during training. This architecture effectively combines traditional U-Net benefits for spatial feature preservation with transformer capabilities for modeling complex channel-environment relationships in the physics-informed framework.
The implementation details and exact dimensions for each part of the network are outlined in Section~\ref{imp}. We also employ a physically informed loss function to combine the physical information from solving Maxwell equations using numerical methods with the estimated channel and reconstruction loss. 
\begin{equation}\label{eq:loss}
    \mathcal{L}_{\text{total}} = \mathcal{L}_{\text{NMSE}} + \zeta \: \mathcal{L}_{\text{phy}}\: , 
\end{equation}
where $ \mathcal{L}_{\text{NMSE}}$ is the reconstruction loss for the channel estimation defined as
\begin{equation}
    \mathcal{L}_{\text{NMSE}} = \mathbb{E} \left( \frac{\|\mathbf{H - \hat{H}}\|_2^2}{\|\mathbf{H}\|_2^2}\right)
\end{equation}
where $\mathbf{H} \in \mathbb{C}^{D \times N_{\text{r}} \times N_{\text{t}}}$ and $\mathbf{\hat{H}} \in \mathbb{C}^{D \times N_{\text{r}} \times N_{\text{t}}}$ are the 3-dimensional complex tensors for true and estimated channels respectively. Moreover, $\mathcal{L}_{\text{phy}}$ is defined as
\begin{align}
\mathcal{L}_{\text{phy}}
&= \mathbb{E}\Bigg(
    \mathbb{E}\!\left[
        \frac{\lambda^{2}}{8\pi\eta_{0}}
        \left|\sum_{i=1}^{P} E_i(\mathbf r)\right|^{2}
    \right] \nonumber \\[0.5ex]
&\qquad -\, P_T
    \sum_{d=1}^{D}
    \mathbb{E}\!\left[
        \|\mathbf{\hat{H}_d}\|_F^2
    \right]
\Bigg)^{2}.
\label{eq:Lphysical}
\end{align}
and $\zeta$ weights the physical consistency term to prevent physics violations while maintaining reconstruction accuracy. The hyperparameter is chosen such that both loss terms contribute comparable gradient magnitudes during early training phases. This ensures the network learns both accurate channel reconstruction and adherence to electromagnetic propagation principles. To formally establish the connection between electromagnetic theory and our loss function, we note that the received power from Maxwell's equations \eqref{eq:received_power} relates to the channel matrix through the path gains. For the $\ell$-th path, the complex gain satisfies
\begin{equation}
\alpha_\ell = \frac{\sqrt{\eta_0}}{2}\frac{\|\mathbf{E}_\ell\|}{\sqrt{P_T}} e^{j\psi_\ell},
\label{eq:gain_field}
\end{equation}
where $\eta_0 \approx 377\,\Omega$ is the free-space impedance. Summing contributions across all taps and applying Parseval's theorem we get
\begin{equation}
P_R^{\text{Chan}} = P_T \sum_{d=1}^{D} \|\mathbf{H}_d\|_F^2 = \kappa \cdot P_R^{\text{EM}}(\mathbf{r}),
\label{eq:power_consistency}
\end{equation}
where $\kappa$ absorbs physical constants (wavelength, array geometry, pulse shape). Therefore, minimizing $\mathcal{L}_{\text{phy}} = \mathbb{E}[(P_R^{\text{EM}} - \kappa P_R^{\text{Chan}})^2]$ enforces electromagnetic power conservation, constraining solutions to the physically admissible manifold $\mathcal{M}_{\text{phys}} = \{\mathbf{H} : P_R^{\text{Chan}}(\mathbf{H}) \approx P_R^{\text{EM}}\}$. The overall proposed method is summarized in Algorithm~\ref{alg:training}.

\begin{algorithm}[H]
\caption{PINN Channel Estimation}
\begin{algorithmic}[1]
\State \textbf{Input:} Training set $\mathcal{D} = \{(\mathbf{y}_i, \mathbf{H}_i, \text{RSS}_i, \mathbf{r}_i)\}_{i=1}^M$, physics weight $\zeta$, epochs $E$.
\State \textbf{Output:} Trained parameters $\theta^*$

\For{epoch $e = 1$ to $E$}
        \State Compute initial channel estimate $\tilde{\mathbf{H}} \gets \text{LS}(\mathbf{y}, N_p)$
        \State Extract RSS features: $\mathbf{F}_{\text{RSS}} \gets \text{RSSEncoder}_\theta(\text{RSS})$
        \State Extract channel features: $\mathbf{F}_{\text{Channel}} \gets \text{Encoder}_\theta(\tilde{\mathbf{H}})$
        \State Fuse via cross-attention:
        \State \hspace{\algorithmicindent} $\mathbf{Q} = \mathbf{W}_{\text{Channel}}\mathbf{F}_{\text{Channel}}$, $\mathbf{K} = \mathbf{W}_{\text{RSS}}\mathbf{F}_{\text{RSS}}$, $\mathbf{V} = \mathbf{W}_{\text{RSS}}\mathbf{F}_{\text{RSS}}$
        \State \hspace{\algorithmicindent} $\mathbf{Z} \gets  \text{softmax}\left(\frac{\mathbf{Q}\mathbf{K}^T}{\sqrt{D_z}}\right)\mathbf{V}$
        \State Refine with transformer: $\mathbf{Z} \gets \text{Transformer}_\theta(\mathbf{Z})$
        \State Reconstruct channel: $\hat{\mathbf{H}} \gets \text{Decoder}_\theta(\mathbf{Z})$
        \State Compute total loss:
        \State \hspace{\algorithmicindent} $\mathcal{L}_{\text{total}} = \mathcal{L}_{\text{NMSE}} + \zeta\mathcal{L}_{\text{phy}}$
        \State Minimize $\mathcal{L}_{\text{total}}$ with respect to $\theta$
\EndFor
\State \textbf{Return:} $\theta^*$
\end{algorithmic}
\label{alg:training}
\end{algorithm}

\begin{remark}
In the pilot-limited regime, the benefit of including the physics-informed term
$\mathcal{L}_{\text{phy}}$ can be understood through a simple bias--variance
argument \cite{bishop2006pattern}. Without constraints, a data-driven estimator must search over a
high-dimensional hypothesis space, leading to variance scaling as
$\operatorname{Var}[f_\theta] \propto d_{\text{eff}}/M$, where $f_\theta$ is the neural estimator with parameters $\theta$, $d_{\text{eff}}$ is the effective model dimension (less than the model dimension $d$) and $M$ the number of pilots. By enforcing approximate physical
constraints $\mathcal{P}(f_\theta(\mathbf{y})) \approx 0$, with $\by$ the observations (pilots and RSS), the search space is
restricted to a lower-dimensional manifold, yielding
\begin{equation}
	\operatorname{Var}[f_\theta] \propto d_{\text{eff}}/M, 
	\qquad d_{\text{eff}} \ll d,
\end{equation}
thus reducing estimation variance with only a minor bias if the physical model
is imperfect. From an optimization viewpoint, the additional quadratic penalty
$\lambda\mathcal{L}_{\text{phy}}$ improves the conditioning of the loss
landscape. Writing a gradient step as
\begin{equation}
	\theta_{t+1} = \theta_t - \eta\bigl(\nabla \mathcal{L}_{\text{data}} +
	\lambda \nabla \mathcal{L}_{\text{phy}} + \xi_t\bigr),
\end{equation}
where $\xi_t$ denotes stochastic noise, the Hessian gains an extra
regularizing term, which lowers the effective condition number and accelerates
convergence. Hence, the physics-informed loss improves both generalization and
training stability.
\end{remark}
\subsection{Multi-Step Temporal Channel Estimation} \label{track}
\begin{figure}[!h]
    \centering
    \includegraphics[width=.85\linewidth]{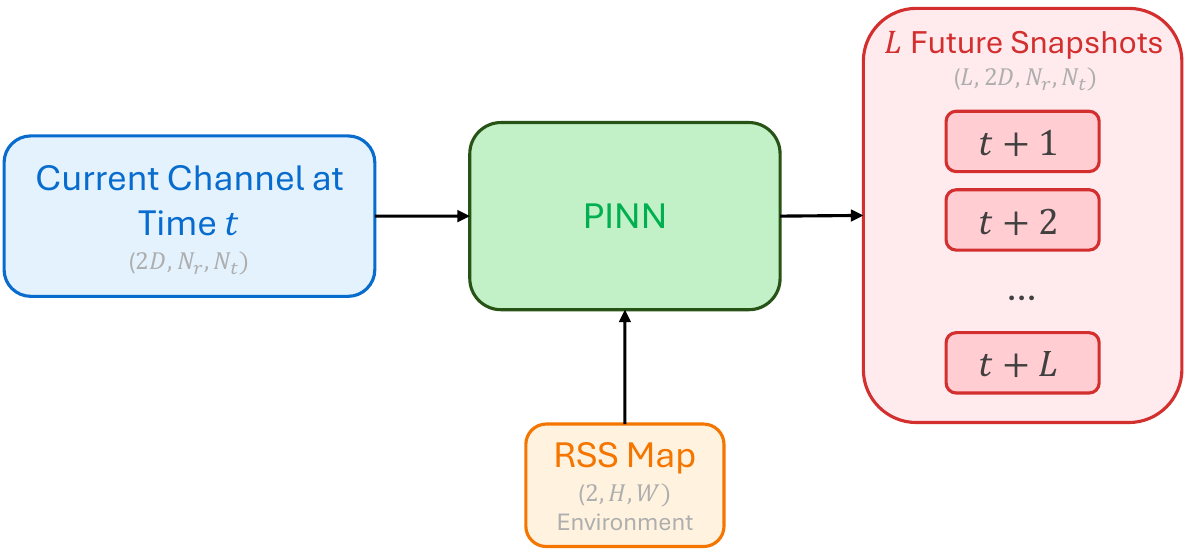}
    \caption{Multi-step temporal channel estimation architecture. The decoder generates $L$ consecutive future channel snapshots in parallel from the current channel estimate and RSS map.}
    \label{fig:diag}
\end{figure}

The framework described thus far addresses statistical channel estimation for a single time instant. However, in highly mobile urban scenarios, the estimated channel may become outdated during the estimation process itself due to rapid temporal variations. In such cases, estimating multiple future channel states becomes essential for effective communication system operation.
To incorporate multi-step temporal estimation capabilities while preserving the computational efficiency and physical interpretability of the proposed architecture, we extend only the decoder portion of the PINN framework. Specifically, we replace the single-output decoder with a multi-step temporal decoder that generates $L$ consecutive future channel snapshots from a single current estimate. Figure~\ref{fig:diag} illustrates this architectural extension.
The decoder achieves this through $L$ specialized prediction heads that operate in parallel, each learning to map the shared latent representation to a specific future time step. This design enables simultaneous generation of all $L$ predictions in a single forward pass, avoiding the computational overhead and error accumulation associated with autoregressive approaches. The loss function remains structurally identical to Equation~\ref{eq:loss}, but now operates over all $L$ predicted time steps. This formulation ensures that all future predictions remain physically consistent with the environmental constraints encoded in the RSS map.
More details on the multi-step estimation and the dataset for the evaluation are discussed in Section~\ref{emptrack}. 
\section{Dataset} \label{data}
\begin{figure}[!h]
    \centering
    \includegraphics[width=.7\linewidth]{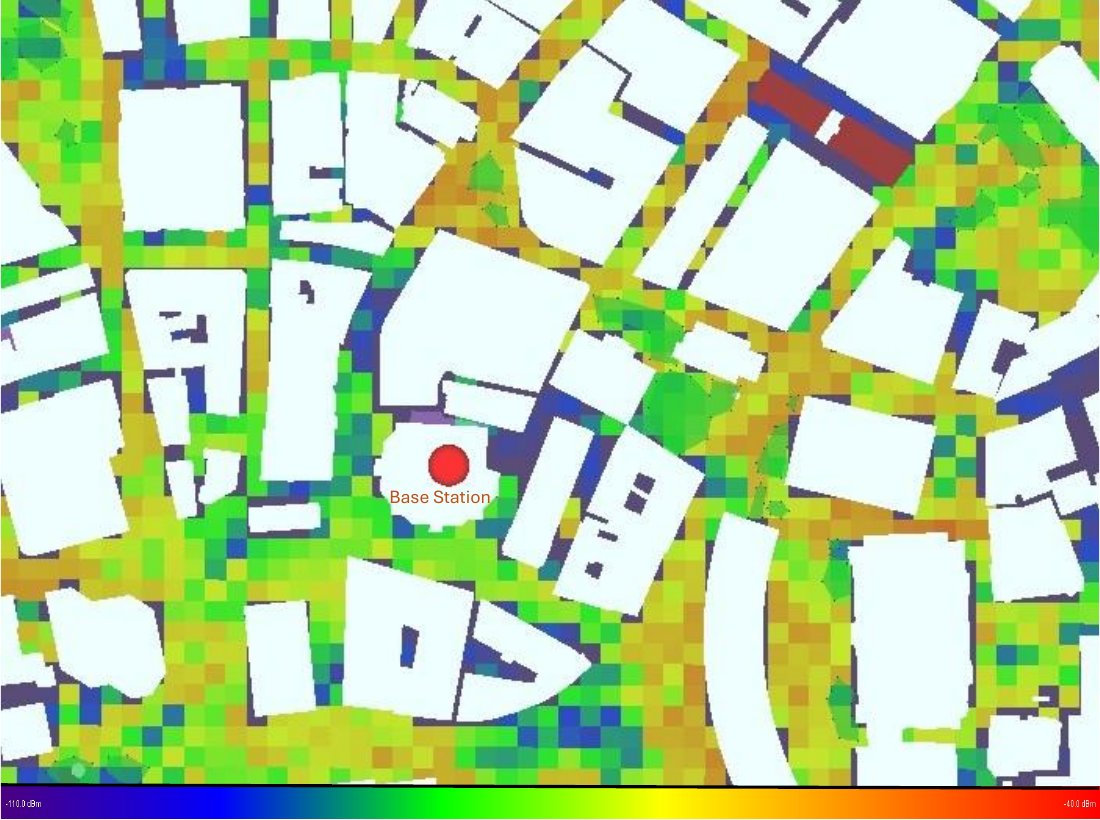}
    \caption{RSS map for the Boston environment with $P_T = 50$ dBm and $f_c = 15$ GHz.}
    \label{fig:rssmap}
\end{figure}

\begin{figure}[!h]
    \centering
    \includegraphics[width=.5\linewidth]{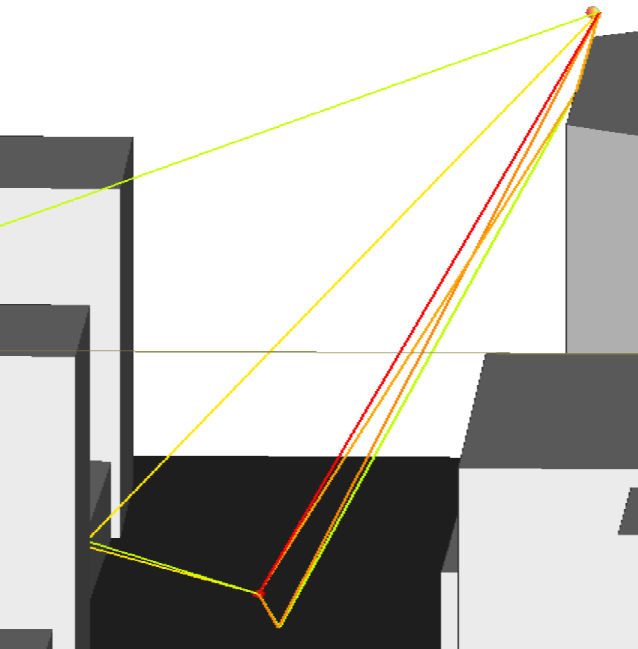}
    \caption{Urban canyon environment with $f_c = 15$ GHz. The \ac{MPC} with gains above $ -120$ dBm are plotted.}
    \label{fig:canyon}
\end{figure}

For more realistic simulations of an urban environment in the upper-mid band, we used \textit{Wireless Insite} \cite{remcom2022wireless} for generating communication channels. This ray tracing tool takes into account the geographical and morphological features of the propagation environment.  It simulates the behavior of each \ac{MPC} between the transmitter and receiver by following physical principles, including free-space power loss and the interaction with various objects. This enables us to compute information for each \ac{MPC}, including complex amplitude, directions of departure and arrival, and delay.
\begin{table*}[htbp]
\centering
\caption{Physics-Informed U-Net architecture and parameters. $\downarrow$ and $\uparrow$ represent the downsampling and upsampling layers, respectively.}
\label{tab:architecture}
\begin{adjustbox}{max width=\textwidth}
\begin{tabular}{ccc|ccc|ccc}
\hline
\multicolumn{3}{c|}{\textbf{Encoder}} & \multicolumn{3}{c|}{\textbf{Latent}} & \multicolumn{3}{c}{\textbf{Decoder}} \\
\hline
\textbf{\#} & \textbf{Type} & \textbf{Output Size} &
\textbf{\#} & \textbf{Type} & \textbf{Output Size} &
\textbf{\#} & \textbf{Type} & \textbf{Output Size} \\
\hline
Input                   & Channel         & $32\times4\times576$ &
1                       & RSS Encoder    & $256$ &
Output                  & Channel        & $32\times4\times576$ \\
1\,($\downarrow$)       & ResUNetBlock    & $64\times2\times288$ &
2                       & Cross-Attention& $18\,432$ &
1\,($\uparrow$)         & ResUNetBlock   & $(128+128)\times1\times144$ \\
2\,($\downarrow$)       & ResUNetBlock    & $128\times1\times144$ &
3                       & Transformer    & $72\times256$ &
2\,($\uparrow$)         & ResUNetBlock   & $(64+64)\times2\times288$ \\
3\,($\downarrow$)       & ResUNetBlock    & $256\times1\times72$ &
                        &                & &
3\,($\uparrow$)         & ResUNetBlock   & $32\times4\times576$ \\
\hline
\end{tabular}
\end{adjustbox}
\end{table*}
In this setup, as we can see in Figure~\ref{fig:rssmap}, the BS is deployed at the top of a tall building for more coverage and according to tilting. This will also help us to avoid the problems that might be caused by near-field estimation. We mainly used the Boston map \cite{remcomFDMIMO2} in this software for simulations. The more details are mentioned in Table~\ref {tab:data}. 
\begin{table}[ht]
\centering
\caption{System parameters in Boston map}
\label{tab:data}
\begin{tabular}{@{} l l}
\toprule
\textbf{Parameter} & \textbf{Value} \\ 
\midrule
Map scale ($\text{m}^2$)               & $350\times450$          \\ 
Cropped map scale ($\text{m}^2$)       & $10\times10$  \\ 
Carrier frequency (GHz)   & 15 and 8        \\ 
Bandwidth (MHz)   & 400 and 200     \\
Number of samples & 9877\\
\hline
\end{tabular}
\end{table}
After extracting the paths for each snapshot, the channels are created using \eqref{channel}. We considered a raised-cosine filter with a rolloff factor of 0.4 as the pulse shaping function. We also used \textit{Wireless Insite} for creating the \ac{RSS} map. Figure~\ref{fig:rssmap} shows the result of solving the Maxwell equations using the described method. 
We used the location of the users at each snapshot with a horizontal Gaussian random noise of $\mathcal{N}(0,\,9 \mathbf{I})$. This assumption is valid based on the average error of the \ac{GPS} \cite{specht2022experimental}. To provide more experimental results across different scenarios, we also used an urban canyon environment \cite{remcomRayBased}, as shown in Figure~\ref{fig:canyon}. The urban canyon map is usually considered a simpler map with fewer details compared to the Boston environment. In general, ray-tracing tools such as \textit{Wireless InSite} provide accurate approximations of real-world propagation, as they explicitly model electromagnetic interactions with buildings, streets, and other objects. Their ability to capture multipath, diffraction, and shadowing makes them widely accepted for validating new channel estimation methods. As shown in prior works \cite{mededjovic2012wireless, remcomUrbanEnv}, the resulting synthetic channels closely mimic measured data, offering a reliable and reproducible alternative to costly field trials.

\section{Experiments} \label{exps}
The codes  and dataset used to obtain the experimental results presented in this section are provided in \cite{codespinnwce}.
\subsection{Network Implementation} \label{imp}
Figure~\ref{fig:nn} highlights the three core components of our network—encoder, latent domain, and decoder—and illustrates how data flows through each stage. Although the dimensionality of these modules is flexible, Table~\ref{tab:architecture} provides the exact layer sizes and overall structure we adopt in this design.

The ResUNet blocks are a pivotal component within both the encoder and decoder, merging U-Net’s effective feature‐preserving skip connections with ResNet’s residual learning framework \cite{ronneberger2015u, he2016deep}. By combining these two architectures, we achieve superior feature propagation and gradient stability—qualities that are especially advantageous for channel estimation. Table~\ref{tab:resunet_block} lays out the precise implementation details of these blocks in both their downsampling and upsampling configurations.

\begin{table}[htbp]
\centering
\caption{ResUNetBlock architecture details}
\label{tab:resunet_block}
\begin{adjustbox}{max width=.45\textwidth}
\begin{tabular}{ccc|ccc}
\hline
\multicolumn{3}{c|}{\textbf{Downsampling Block}} & \multicolumn{3}{c}{\textbf{Upsampling Block}} \\
\hline
\textbf{\#} & \textbf{Layer} & \textbf{Parameters} & \textbf{\#} & \textbf{Layer} & \textbf{Parameters} \\
\hline
1 & Conv2d & $3 \times 3$, stride=1, pad=1 & 1 & ConvTranspose2d & $3 \times 3$, stride=2, pad=1 \\
2 & GroupNorm & 8 groups & 2 & GroupNorm & 8 groups \\
3 & LeakyReLU & $\alpha = 0.2$ & 3 & ReLU & - \\
4 & Conv2d & $3 \times 3$, stride=2, pad=1 & 4 & Conv2d & $3 \times 3$, stride=1, pad=1 \\
5 & GroupNorm & 8 groups & 5 & GroupNorm & 8 groups \\
6 & LeakyReLU & $\alpha = 0.2$ & 6 & ReLU & - \\
\hline
\multicolumn{3}{c|}{\textbf{Residual Connection}} & \multicolumn{3}{c}{\textbf{Residual Connection}} \\
\hline
1 & Conv2d & $1 \times 1$, stride=2, pad=0 & 1 & ConvTranspose2d & $1 \times 1$, stride=2, pad=0 \\
2 & GroupNorm & 8 groups & 2 & GroupNorm & 8 groups \\
\hline
\end{tabular}
\end{adjustbox}
\end{table}
\begin{table*}[htbp]
	\centering
	\caption{Training hyperparameters}
	\label{tab:training_params}
	\begin{adjustbox}{max width=\textwidth}
		\begin{tabular}{lcccccccc}
			\toprule
			\textbf{Batch Size} & \textbf{\# Epochs} & \textbf{Init.\ LR} & \textbf{Scheduler} & \textbf{Decay Step} & \boldmath$\gamma$ \textbf{(Decay)} & \textbf{Optimizer} & \textbf{Momentum} & \boldmath$\zeta$\\
			\midrule
			32                  & 500                & $1\times10^{-3}$   & StepLR             & 40                  & 0.65                             & Adam               & 0.9  & 0.01            \\
			\bottomrule
		\end{tabular}
	\end{adjustbox}
\end{table*} 
To process \ac{RSS} inputs, we employ a compact CNN‐based encoder that efficiently distills spatial \ac{RSS} patterns into a fixed‐length feature vector for downstream channel estimation. As detailed in Table~\ref{tab:rss_encoder}, the encoder begins with a series of $3\times3$ convolutions and ReLU activations interleaved with $2\times2$ max‐pooling to gradually increase channel depth while reducing spatial dimensions. A final adaptive average pooling and flatten operation produces a 256‐dimensional embedding that captures the most salient \ac{RSS} features.

\begin{table}[htbp]
\centering
\caption{RSS encoder architecture details}
\label{tab:rss_encoder}
\begin{adjustbox}{max width=.45\textwidth}
\begin{tabular}{lcc}
\hline
\textbf{Stage} & \textbf{Operations} & \textbf{Output Size} \\
\hline
Input           & RSS Map input                                       & $1\times30\times30$ \\
Block 1         & Conv($3\times3$, 32) → ReLU → MaxPool($2\times2$)                   & $32\times15\times15$ \\
Block 2         & Conv($3\times3$, 64) → ReLU → MaxPool($2\times2$)                   & $64\times7\times7$   \\
Block 3         & Conv($3\times3$, 128) → ReLU → MaxPool($2\times2$)                  & $128\times3\times3$  \\
Block 4         & Conv($3\times3$, 256) → ReLU                                 & $256\times3\times3$  \\
Pooling & AdaptiveAvgPool($1\times1$) → Flatten & $256$              \\
\hline
\end{tabular}
\end{adjustbox}
\end{table}
\subsection{Complexity Analysis}
In this section, we analyze the neural network’s computational requirements and compare it to a diffusion model inspired by \cite{zhou2025generative}, which we have adapted for our dataset and experimental setup as a state-of-the-art method. Table~\ref{tab:pinn-comparison} highlights the key metrics for both architectures.

\begin{table}[htbp]
  \centering
  \caption{Complexity, latency, and parameters comparison}
  \label{tab:pinn-comparison}
  \begin{tabular}{lccc}
    \toprule
    & \multicolumn{1}{c}{\textbf{FLOPs}} 
    & \multicolumn{1}{c}{\textbf{Latency [ms]}} 
    & \multicolumn{1}{c}{\textbf{Parameters}} \\
    \cmidrule(lr){2-2} \cmidrule(lr){3-3} \cmidrule(lr){4-4}
    $(N_t,N_r,D)$ 
      & \((576,4,16)\) 
      & \((576,4,16)\) 
      & \((576,4,16)\) \\
    \midrule
    PINN & 70.85G & 11.12 & $3.5 \times 10^8$  \\
    DM & 130.15G & 50.30 & $5.5 \times 10^4$\\
    \bottomrule
  \end{tabular}

\end{table}

As seen in Table~\ref{tab:pinn-comparison}, our \ac{PINN} requires significantly fewer FLOPs and achieves much lower inference latency, because of the complexity of the backward pass of the diffusion model. The diffusion architecture, however, maintains an extremely low parameter, making it more memory‑efficient. In our design, most parameters reside in large linear and convolutional layers, which can benefit substantially from optimized implementations. Additionally, the use of efficient skip connections enhances our model’s performance without increasing its computational cost.

\subsection{Numerical Results}
\begin{figure}[!h]
	\centering
	\includegraphics[width=.81\linewidth]{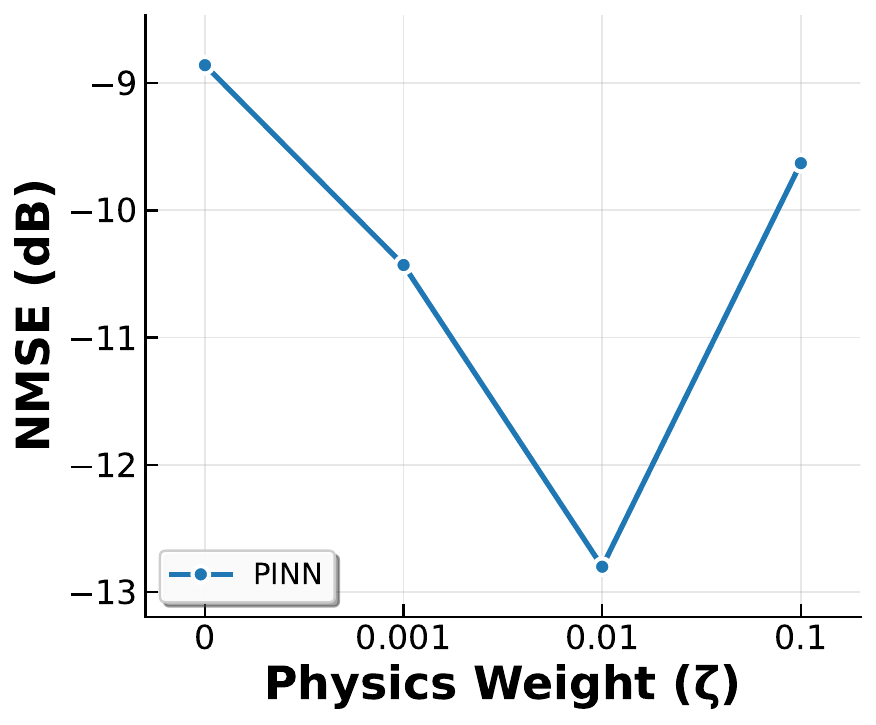}
	\caption{Ablation study of weight for physical component in loss function.}
	\label{fig:ab}
\end{figure}

We begin by conducting an ablation study to determine the optimal weighting parameter $\zeta$ in \eqref{eq:loss}. Figure~\ref{fig:ab} presents the NMSE performance as a function of $\zeta$ for the Boston dataset using LS-OFDM initial estimates with $N_p = 4$ pilots at $SNR = 0$ dB. The results indicate that $\zeta = 0.01$ yields optimal performance, balancing the reconstruction accuracy (NMSE term) with the physics-based power constraint. This value is adopted for all subsequent experiments.

It is important to note that both the initial channel estimates and RSS maps undergo global normalization prior to network input and loss computation by a fixed number (the global maximum value across all the channels). This normalization ensures numerical stability during training and meaningful gradient propagation across different physical quantities.


\begin{figure}[!h]
	\centering
	\includegraphics[width=.81\linewidth]{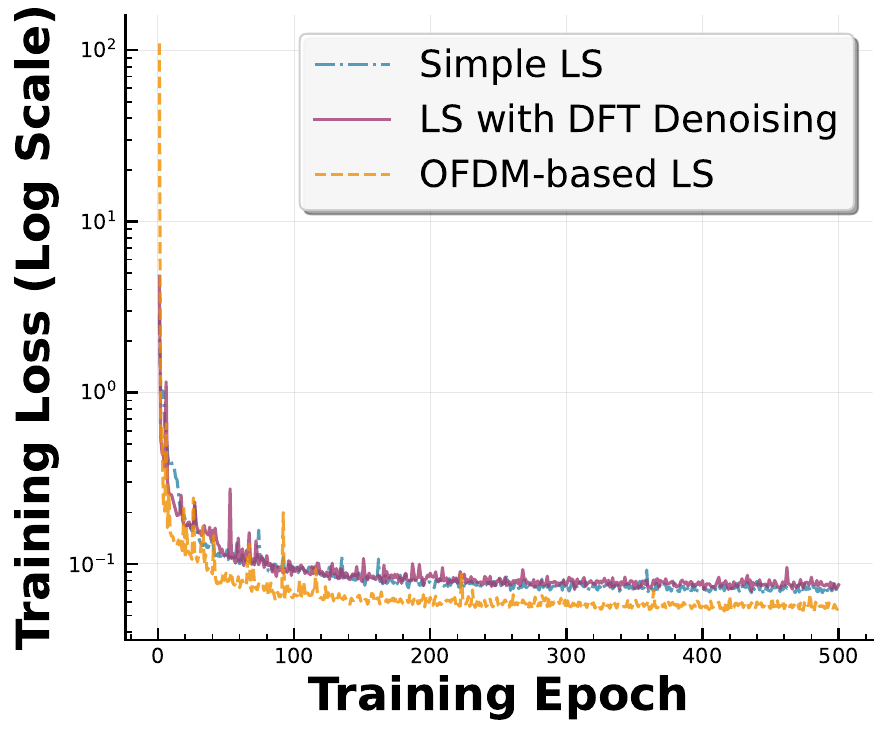}
	\caption{Log–scale comparison of training loss for the selected models.}
	\label{fig:train}
\end{figure}

After defining hyperparameters from Table~\ref{tab:training_params}, we proceed with training. The dataset is partitioned into 80\% for training, 10\% for validation, and 10\% for testing. Prior to training, we normalize every sample by a fixed constant to ensure consistency when minimizing the \ac{NMSE} reconstruction loss. An identical normalization—using the same constant—is applied to the \ac{RSS}‐derived power values so that all inputs and targets share the same scale. Some of the training loss curves are shown in Figure~\ref{fig:train}. All experiments were run on an NVIDIA GeForce RTX 4090 GPU. 
Figure~\ref{fig:train} clearly demonstrates the convergence of this method for different initial estimations. After defining our network architecture, we outline in Table~\ref{tab:training_params} the hyperparameter settings used during training. These choices were determined empirically to balance convergence speed and generalization. 

We categorize our tests into multiple main sections for presentation purposes. In all methods, unless otherwise specified, we used \(N_t = 24 \times 24\), \(N_r = 2 \times 2\), and \(D = 16\). We also modeled the channel at an upper-mid band frequency of \(f_c = 15\) GHz, which holds significant potential for future wireless systems \cite{kang2024cellular}. We employ standard uniform pilot spacing to ensure fair comparison with baseline methods. Joint optimization of pilot placement and PINN refinement using RSS side information remains an interesting direction for future work. In the first set, we investigated the performance of less complex spatial simple \ac{LS} channel estimation using interpolation and \ac{DFT}-based denoising. We consider different \ac{SNR} values while setting the number of pilots as $N_p = 64$ ($\frac{N_p}{N_t} = 0.11$).
\begin{figure}[!h]
    \centering
    \begin{subfigure}[b]{0.81\linewidth}
        \centering
        \includegraphics[width=\linewidth]{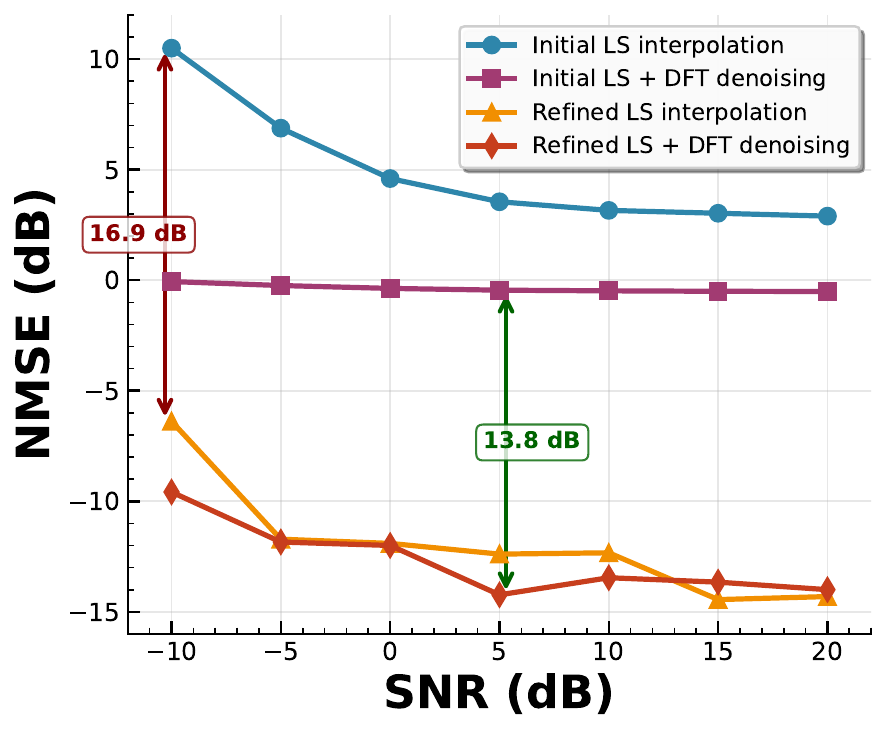}
        \caption{NMSE vs.\ SNR }
        \label{fig:per1}
    \end{subfigure}
    \vspace{1em}
    \begin{subfigure}[b]{0.81\linewidth}
        \centering
        \includegraphics[width=\linewidth]{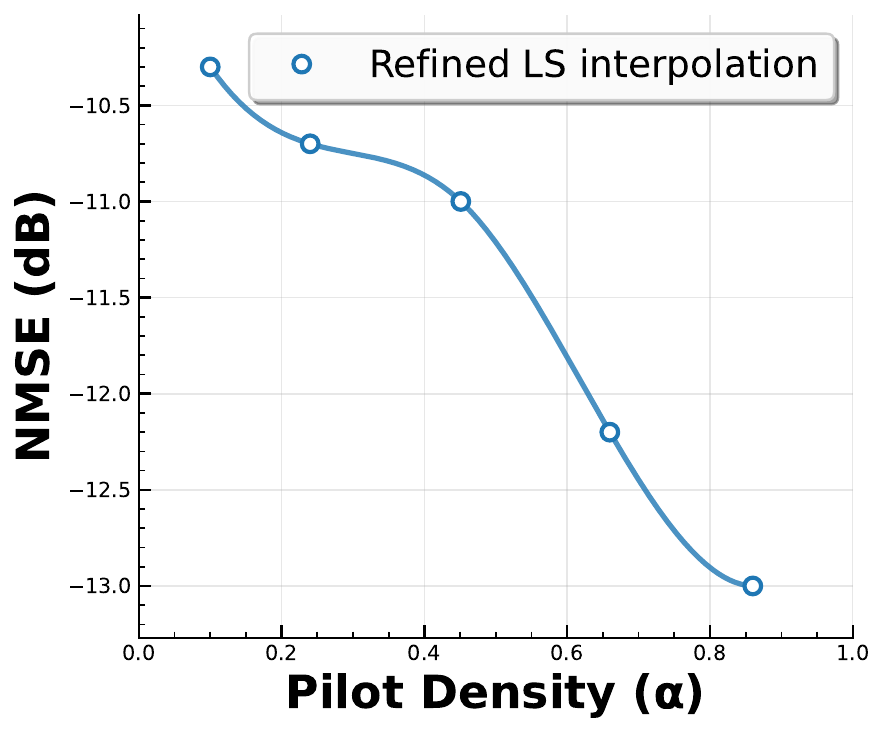}
        \caption{NMSE vs.\ pilot density}
        \label{fig:per5}
    \end{subfigure}
    \caption{NMSE performance of channel estimation using less complex initial estimation}
    \label{fig:combined_nmse2}
\end{figure}
\begin{table*}[!h]
\centering
\caption{NMSE performance of channel estimation with varying $N_p$ at $\mathrm{SNR}=0$ dB.}
\label{tab:performance_comparison}
\begin{tabular}{c|rrrrrrrrr}
\toprule
\textbf{$N_p$} & \textbf{512} & \textbf{256} & \textbf{128} & \textbf{64} & \textbf{32} & \textbf{16} & \textbf{8} & \textbf{4} & \textbf{2} \\
\midrule
Initial LS-OFDM  & -14.40 & -11.23 & -8.25 & -5.49 & -3.32 & -1.33 & 0.28 & 1.98 & 3.01 \\
Refined LS-OFDM  & -19.21 & -17.40 & -16.20 & -16.00 & -13.87 & -13.85 & -12.87 & -12.89 & -12.35 \\
Improvement      & 4.81 & 6.17 & 7.95 & 10.51 & 10.55 & 12.52 & 13.15 & 14.87 & 15.36 \\
\bottomrule
\end{tabular}
\end{table*}
We can see in Figure~\ref{fig:per1}, applying the physics‐informed neural network refinement further lowers \ac{NMSE} to roughly $(-14.45,-6.38)$ dB (interpolation) and $(-14.23,-9.58)$ dB (denoising), corresponding to maximum gains of up to $16.9$ dB and $13.8$ dB respectively.
The refined methods maintain robust performance even at $\mathrm{SNR}=-10$ dB and saturate near $-14$ dB \ac{NMSE} at high \ac{SNR}s, indicating that residual error is dominated by measurement errors that the network cannot correct. Figure~\ref{fig:per5} also indicates the performance change in different pilot densities ($\alpha = \frac{N_p}{N_t}$) at $\mathrm{SNR} = 0$ dB. We can see the robust performance of this method to pilot density. The next set of results uses \ac{OFDM} for channel estimation. We set the number of subcarriers $N=1024$ and change the subcarrier spacing, and accordingly, the number of pilot signals. We also set the $\mathrm{SNR}$ as 0 dB in this simulation. Table~\ref{tab:performance_comparison} demonstrates that the \ac{PINN} estimation shows a great performance in limited pilot scenarios, achieving up to \ac{NMSE} = -12.89 dB with only four pilots. 

\begin{figure}[!h]
    \centering
    \begin{subfigure}[b]{0.81\linewidth}
        \centering
        \includegraphics[width=\linewidth]{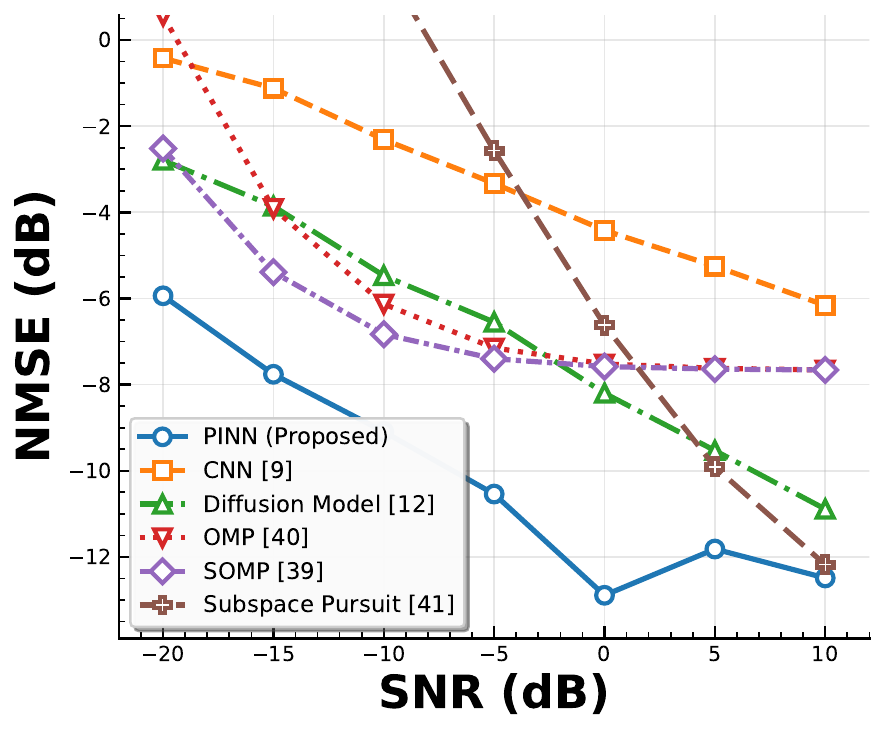}
        \caption{NMSE vs.\ SNR with $N_p = 4$ ($N_p=64$ for classical methods).}
        \label{fig:per2}
    \end{subfigure}
    \vspace{1em}
    \begin{subfigure}[b]{0.81\linewidth}
        \centering
        \includegraphics[width=\linewidth]{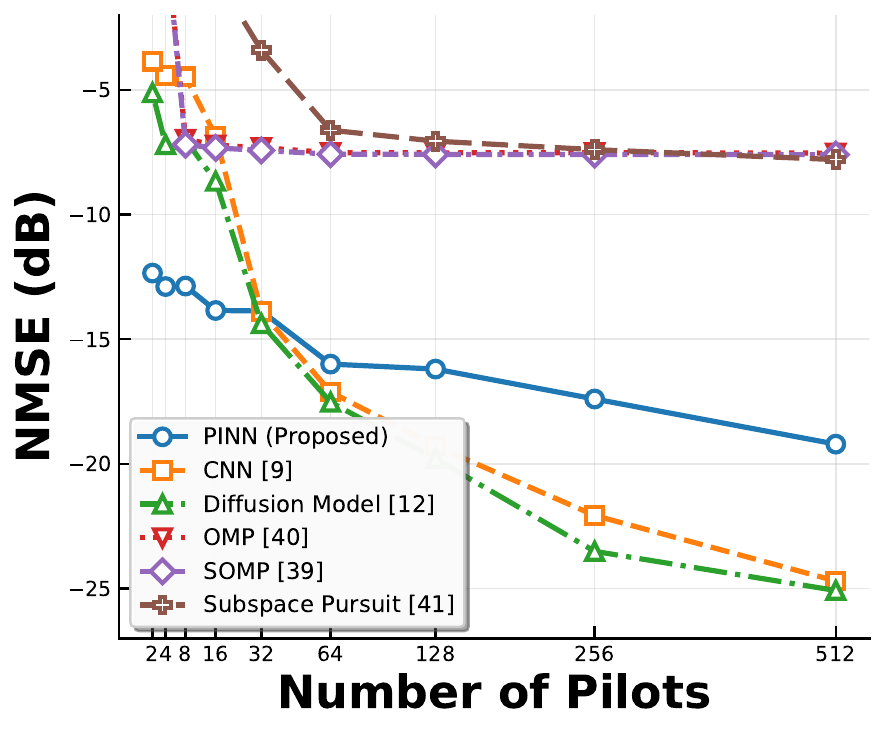}
        \caption{NMSE vs.\ number of pilot signals at $\mathrm{SNR}=0$ dB.}
        \label{fig:per3}
    \end{subfigure}
    \caption{NMSE comparison across varying SNRs and pilot counts.}
    \label{fig:combined_nmse}
\end{figure}

In the next set of experiments, we compare the introduced method with state-of-the-art algorithms in addition to classical \ac{OMP} methods \cite{dai2009subspace, gao2016channel, tropp2007signal}. These greedy compressed sensing algorithms iteratively select angular-domain dictionary atoms to exploit the sparse structure of massive MIMO channels, with simultaneous OMP (SOMP) handling joint multi-vector recovery and Subspace Pursuit using candidate expansion and pruning for improved performance. To have a fair comparison, we only investigate the initial estimation for \ac{LS}-\ac{OFDM}. We set the number of pilots to a limited value of $N_p = 4$ for our proposed method, while for the classical methods we set $N_p = 64$, so we can achieve close performance, to emphasize the power of the introduced method across different $\mathrm{SNR}$s. We also implement a \ac{CNN}-based channel estimation structure inspired by \cite{sattari2024full} (using two hidden layers according to the paper), and a diffusion model-based method from \cite{zhou2025generative} (using the full resolution method). In this experiment, we focused on the low \ac{SNR} region, which is more realistic for the next generation of wireless systems \cite{jin2025near}. Figure~\ref{fig:per2} reveals that the introduced \ac{PINN} improves the performance by a large margin (around 5 dB improvement at $\mathrm{SNR}=0$) in the limited pilot condition. These results highlight the method's robustness when pilot signals are scarce and underscore how incorporating physical environment information significantly enhances the network's channel estimation capabilities.

Figure~\ref{fig:per3} presents a comparative analysis across varying pilot values at $\mathrm{SNR}=0$. The results reveal that \ac{CNN} and diffusion-based methods excel when pilot signals are very abundant, whereas \ac{PINN} achieves superior performance under pilot-limited conditions, often by substantial margins. This dichotomy indicates that \ac{PINN}'s physics-informed architecture, while potentially over-parameterized for pilot-rich scenarios, provides essential inductive bias when pilot overhead is constrained. The physical modeling component effectively compensates for sparse pilot information, validating the value of incorporating domain knowledge into the estimation framework to provide a more robust estimate with respect to pilot density. 
\subsection{Generalization}
\begin{figure}[!h]
    \centering
    \begin{subfigure}[b]{0.81\linewidth}
        \centering
        \includegraphics[width=\linewidth]{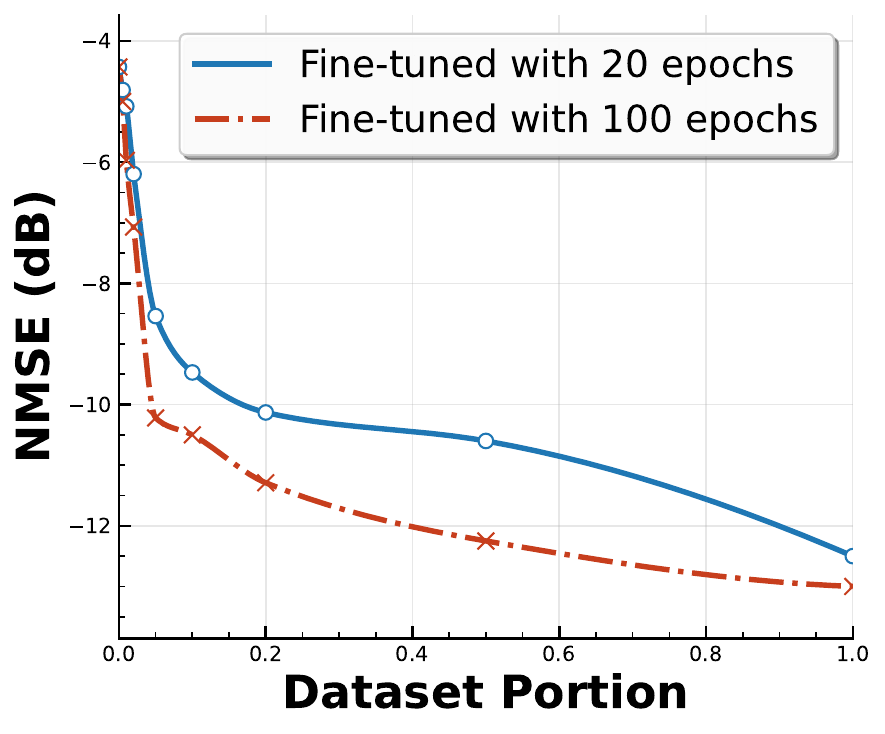} 
        \caption{From 15 GHz to 8 GHz in Boston environment}
        \label{fig:8GHZ}
    \end{subfigure}
    \vspace{1em}
    \begin{subfigure}[b]{0.81\linewidth}
        \centering
        \includegraphics[width=\linewidth]{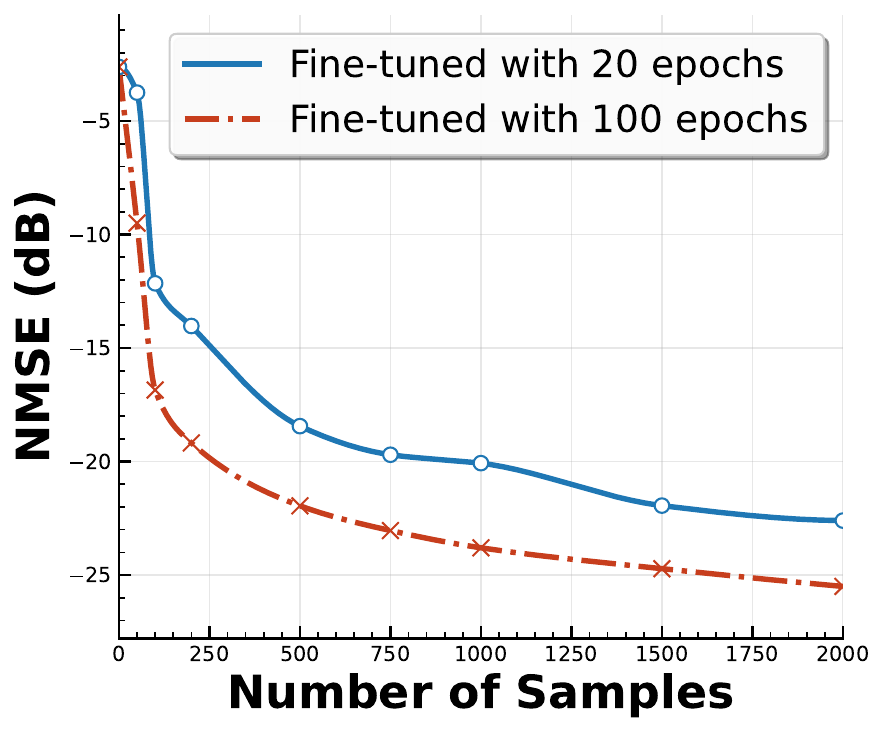}
        \caption{From Boston to urban canyon environment}
        \label{fig:canyon1}
    \end{subfigure}
    \caption{Transfer learning performance: NMSE vs. dataset for 20 and 100 epoch fine-tuning using $N_p=4$ and $\mathrm{SNR}=0$.}
    \label{fig:combinedse}
\end{figure}

To demonstrate the generalization capability of our physics-informed approach, we conducted transfer learning experiments from 15 GHz to 8 GHz, using initial LS-OFDM estimation and four pilot signals at $\mathrm{SNR}=0$ dB. We also changed the bandwidth for the new frequency band as $200$ MHz. The model was initially trained on the 15 GHz dataset and then fine-tuned on 8 GHz data with varying amounts of training samples and different fine-tuning durations. Figure~\ref{fig:8GHZ} shows the performance across different dataset portions. The results show that even with only 10\% of the 8 GHz training data, the fine-tuned model achieves \ac{NMSE} values of approximately $-4.5$ dB to $-5.0$ dB. Performance improves consistently as more data is used, reaching around $-13$ dB with the full dataset which aligns with the experiments for 15 GHz. 
To further validate the adaptability of our physics-informed approach, we conducted transfer learning experiments from the Boston environment to an urban canyon scenario. Figure~\ref{fig:canyon1} shows the NMSE performance as a function of the number of training samples for both 20 and 100 epoch fine-tuning.

The results demonstrate rapid adaptation with limited data: using only 100 samples, the model achieves NMSE values of approximately $-4$ dB (20 epochs) and $-9$ dB (100 epochs). Performance continues to improve substantially with more samples, reaching $-22$ dB and $-25$ dB respectively with 2000 samples. This confirms that our PINN framework effectively captures environment-agnostic electromagnetic principles, enabling efficient adaptation to different urban propagation scenarios with minimal training. We can also conclude that when the network is trained on a more complex dataset, like the Boston map, it learns generalizable electromagnetic propagation principles that enable effective adaptation to different propagation scenarios, such as urban canyons, with minimal additional training data, achieving a better performance than the original dataset.
\subsection{Multi-Step Estimation} \label{emptrack}
So far, we have only analyzed the static channel estimation without considering mobility. As we discussed in Section~\ref{track}, to have an adaptable method for mobile cases, we changed the last layer of the PINN architecture with an $L$ snapshot estimation decoder. The dataset that we used for this stage is a multi-trajectory dataset in the Boston environment, and a single trajectory in the urban canyon environment. In mobile scenarios, due to the Doppler effect in time-varying channels, the estimated channel for the future time slots could be significantly different from the acquired channel. Channel coherence time \cite{goldsmith2005wireless} is an estimation that can describe the significance of the Doppler effect in a time-varying channel. The faster the user speed, the stronger the Doppler effect, and the smaller the channel coherence time. The channel coherence time $T_c$ is defined as the time during which the channel can be reasonably well viewed as time-invariant, which is inversely proportional to the frequency and the user speed, i.e.,
\begin{equation} \label{eq:coh}
    T_c \approx \frac{0.5 c}{ v f_c},
\end{equation}
where $v$ is the user speed and $f_c$ is the carrier frequency. In the following experiments, we considered each time interval to be slightly larger than the coherence time~\eqref{eq:coh} for the trajectories. For the considered urban mobility scenarios with velocities of 35-39 km/h at 15 GHz carrier frequency, the channel coherence time is approximately $T_c \approx 1.0$ ms. This resolution ensures adequate sampling of channel variations while maintaining computational tractability for real-time estimation. This sampling strategy balances the need to capture channel dynamics with computational 
efficiency for real-time implementation.

Focusing on the model modification, the $L$ future snapshots are generated through a parallel multi-head decoder architecture that expands the output channel dimension by a factor of $L$. The decoder maintains the standard U-Net structure with three upsampling blocks, residual connections, and skip pathways from the encoder. The critical modification occurs in the final decoder block, where the output convolutional layer produces $2D \times L$ channels representing all $L$ future predictions simultaneously. A subsequent reshaping operation reorganizes these outputs from the flattened representation into the structured temporal sequence. During the training for multi-step estimation, the transformer layers in the PINN, specifically the self-attention, capture the time-dependent feature considering the estimated future steps. Through this, the architecture is adapted for this purpose. 

\begin{figure}[!h]
	\centering
	\includegraphics[width=.43\textwidth]{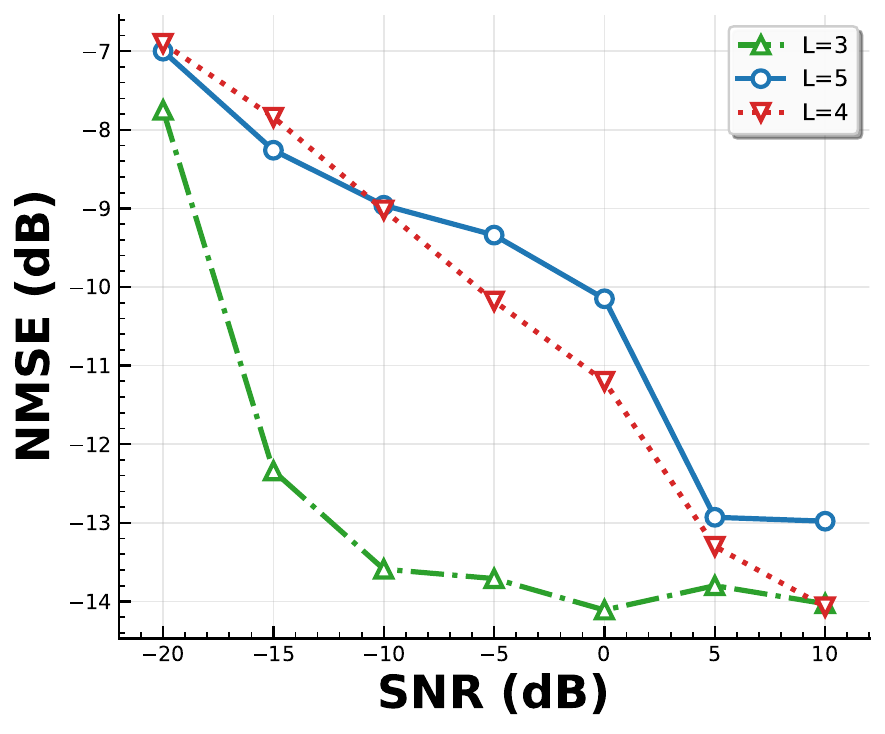}
	\caption{NMSE performance versus SNR for different prediction horizons ($L=3,4,5$) in the urban canyon environment with average velocity of 35 km/h and $N_p=16$ pilots.}
	\label{fig:diffsnrl}
\end{figure}

We start with a multi-step estimation for two different environments. In the Boston environment, the average velocity is 39 km/h, and in an urban canyon, it is 35 km/h. The SNR in both cases is $0$ dB, and the number of pilot signals is $N_p = 4$. 
\begin{figure}[!h]
	\centering
	\includegraphics[width=.43\textwidth]{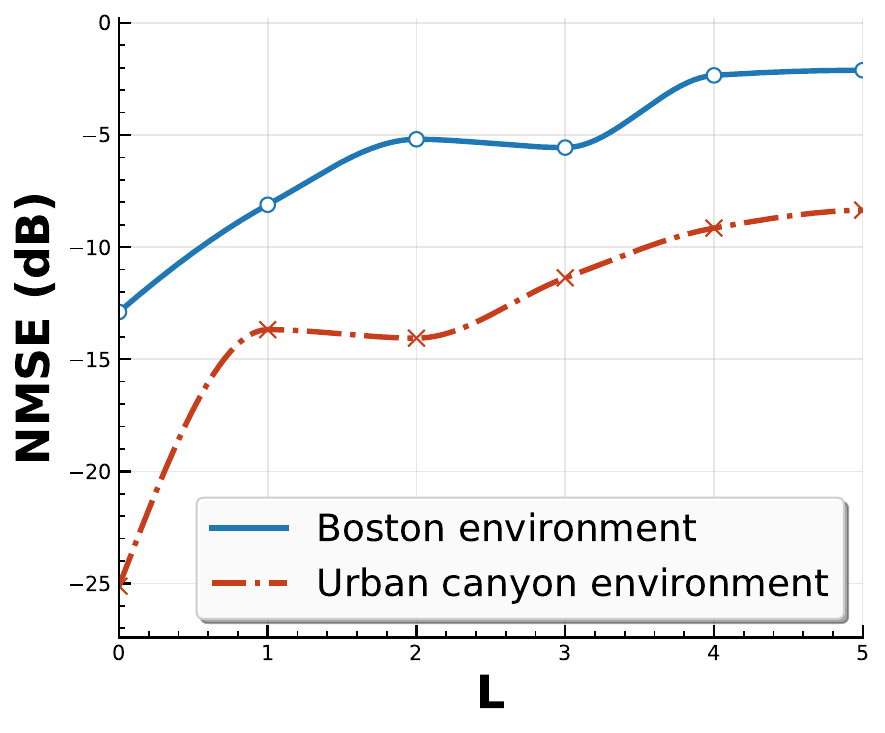}
	\caption{NMSE performance as a function of prediction horizon $L$ for multi-step temporal channel estimation in Boston and urban canyon environments at $0$ dB SNR with $N_p = 4$ pilots.}
	\label{fig:diffl}
\end{figure}
As Figure~\ref{fig:diffl} demonstrates, the NMSE performance degrades gracefully as the prediction horizon $ L$ increases, reflecting the inherent uncertainty in forecasting further into the future. For the urban canyon environment, the network started with $-13.5$ dB for $L=1$ and degraded to approximately $-8.5$ dB at $L=5$. The Boston environment shows a similar trend but with higher overall NMSE values, starting at approximately $-9$ dB for $L=1$ and degrading to around $-2$ dB at $L=5$. This performance difference can be attributed to the higher complexity and richer scattering environment in Boston, which has been explained in Section~\ref{data}. We should also note that the velocity difference and the Doppler effect can degrade the system performance. 
In the last experiment, we analyzed the NMSE performance in different SNRs. We also consider different values of $L$ to investigate performance degradation with the number of different future steps. To better analyze the impact of SNR and prediction horizon $L$ on performance, we fix the number of pilots at $N_p = 16$ for this experiment. We used the urban canyon environment with an average velocity of 35 km/h for this experiment. Figure~\ref{fig:diffsnrl} depicts the \ac{NMSE} performance of the multi-step estimation for different values of $L$ in different \ac{SNR}s. 
As expected, performance degrades with increasing prediction horizon $L$, reflecting the growing uncertainty in forecasting further into the future. The network demonstrates robust performance across all SNR conditions, achieving approximately $-13$ dB NMSE for $L=5$ at SNR $= 5$ dB. Even at lower SNRs, the framework maintains strong estimation accuracy, particularly for $L=3$, which achieves around $-14$ dB NMSE across SNR $\geq -5$ dB. Another observation is the performance convergence at high SNR (above $0$ dB), where all three prediction horizons ($L=3,4,5$) achieve similar NMSE values. This convergence suggests that the network's temporal prediction capability is primarily limited by inherent channel dynamics rather than estimation noise when initial estimates are sufficiently accurate. Conversely, at lower SNRs (below $-10$ dB), the performance gap between different $L$ values widens, indicating that prediction uncertainty compounds under noisy initial conditions. These results confirm that the proposed framework effectively balances near-term and far-term prediction accuracy while remaining resilient to varying noise conditions.

\section{Conclusion} \label{con}
This paper introduced a \ac{PINN} for accurate wireless channel estimation under pilot-constrained scenarios. By integrating initial simple least-squares estimates with \ac{RSS} maps derived from Maxwell-based ray tracing, the proposed architecture leverages environmental knowledge through a U-Net backbone enhanced with transformer and cross-attention modules. Our results on realistic 15 GHz urban ray-traced data show that \ac{PINN} significantly outperforms conventional and learning-based baselines, especially in low-pilot and low-SNR regimes, achieving up to 15 dB \ac{NMSE} gain over initial \ac{LS} estimates and over 4 dB gain compared to state-of-the-art models (reaching around -13 dB NMSE with only four pilots at \ac{SNR} = 0 dB). Moreover, we investigated the generalization capability of this method in different frequency bands and environments. At the end, we extended the framework to mobile scenarios with multi-step CSI estimation, reducing re-pilot overhead by forecasting several future snapshots from one coarse estimate while preserving physical consistency. With low inference latency and more interpretability, \ac{PINN} offers a scalable and physically grounded solution for real-time deployment in next-generation \ac{MIMO} systems.

	\bibliographystyle{IEEEtran}
	\bibliography{refs}
\end{document}

%% file: macros.tex
\usepackage{amsmath,amssymb}

\newcommand{\bbC}{{\mathbb{C}}}

\newcommand{\bbR}{{\mathbb{R}}}

\newcommand{\by}{{\mathbf{y}}}

\newcommand{\bX}{{\mathbf{X}}}
\newcommand{\bY}{{\mathbf{Y}}}

\makeatletter
\def\munderbar#1{\underline{\sbox\tw@{$#1$}\dp\tw@\z@\box\tw@}}
\makeatother

%% file: acronyms.tex
\usepackage{acro}
\DeclareAcronym{3GPP}{
	short=3GPP,
	long=3rd Generation Partnership Project
}
\DeclareAcronym{ADC}{
	short=ADC,
	long=analog-to-digital converter
}
\DeclareAcronym{ADP}{
	short=ADP,
	long=angular-delay profile
}
\DeclareAcronym{AMP}{
	short=AMP,
	long=approximate message passing
}
\DeclareAcronym{AoA}{
	short=AoA,
	long=angle-of-arrival
}
\DeclareAcronym{AoD}{
	short=AoD,
	long=angle-of-departure
}
\DeclareAcronym{APS}{
	short=APS,
	long=azimuth power spectrum
}
\DeclareAcronym{AWGN}{
	short=AWGN,
	long=additive white Gaussian noise
}
\DeclareAcronym{AV}{
	short=AV,
	long=autonomous vehicle
}
\DeclareAcronym{BI}{
	short=BI,
	long=Bayesian inference
}
\DeclareAcronym{BS}{
	short=BS,
	long=base station
}
\DeclareAcronym{BSM}{
	short=BSM,
	long=basic safety message
}
\DeclareAcronym{CDF}{
	short=CDF,
	long=cumulative distribution function
}
\DeclareAcronym{CIR}{
	short=CIR,
	long=channel impulse response
}
\DeclareAcronym{CNN}{
	short=CNN,
	long=convolutional neural network
}
\DeclareAcronym{ChanSTA}{
	short=ChanSTA,
	long=channel spatial and temporal attention
}

\DeclareAcronym{CP}{
	short=CP,
	long=cyclic-prefix
}

\DeclareAcronym{CS}{
	short=CS,
	long=compressed sensing
}
\DeclareAcronym{CSI}{
	short=CSI,
	long=channel state information
}

\DeclareAcronym{DFT}{
	short=DFT,
	long=discrete Fourier transform
}
\DeclareAcronym{IFFT}{
	short=IFFT,
	long=inverse fast Fourier transform
}
\DeclareAcronym{FFT}{
	short=FFT,
	long=fast Fourier transform
}
\DeclareAcronym{DL}{
	short=DL,
	long=deep learning
}

\DeclareAcronym{DNN}{
	short=DNN,
	long=deep neural network
}
\DeclareAcronym{DoA}{
	short=DoA,
	long=direction-of-arrival
}
\DeclareAcronym{DoD}{
	short=DoD,
	long=direction-of-departure
}
\DeclareAcronym{DSRC}{
	short=DSRC,
	long=dedicated short-range communication
}
\DeclareAcronym{EM}{
	short=EM,
	long=expectation maximization
}
\DeclareAcronym{FC}{
	short=FC,
	long=fully connected
}

\DeclareAcronym{FDD}{
	short=FDD,
	long=frequency division duplex
}
\DeclareAcronym{FMCW}{
	short=FMCW,
	long=frequency modulated continuous wave
}
\DeclareAcronym{FoV}{
	short=FoV,
	long=field-of-view
}
\DeclareAcronym{GNSS}{
	short=GNSS,
	long=global navigation satellite system
}
\DeclareAcronym{GPS}{
	short=GPS,
	long=global positioning system
}
\DeclareAcronym{IoT}{
	short=IoT,
	long=internet of things
}
\DeclareAcronym{IMU}{
	short=IMU,
	long=inertial measurement unit 
}
\DeclareAcronym{KL}{
	short=KL,
	long=Kullback–Leibler
}
\DeclareAcronym{KF}{
	short=KF,
	long=Kalman filter
}
\DeclareAcronym{LIDAR}{
	short=LIDAR,
	long=light detection and ranging
}
\DeclareAcronym{LOS}{
	short=LOS,
	long=line-of-sight
}
\DeclareAcronym{LPF}{
	short=LPF,
	long=low pass filter
}
\DeclareAcronym{LTE}{
	short=LTE,
	long=long term evolution
}
\DeclareAcronym{LS}{
	short=LS,
	long=least squares
}
\DeclareAcronym{LSTM}{
	short=LSTM,
	long=long short-term memory
}
\DeclareAcronym{mmWave}
{
	short = mmWave, 
	long = millimeter wave
}
\DeclareAcronym{MOMP}{
	short=MOMP,
	long=multidimensional orthogonal matching pursuit
}
\DeclareAcronym{MUSIC}{
	short=MUSIC,
	long=multiple signal classification
}
\DeclareAcronym{MIMO}{
	short=MIMO,
	long=multiple-input multiple-output
}
\DeclareAcronym{MHA}{
	short=MHA,
	long=multi-head attention
}
\DeclareAcronym{ML}{
	short=ML,
	long=machine learning
}
\DeclareAcronym{MLE}{
	short=MLE,
	long=maximum likelihood estimation
}
\DeclareAcronym{MLP}{
	short=MLP,
	long=multilayer perceptron
}
\DeclareAcronym{MRR}{
	short=MRR,
	long=medium range radar
}
\DeclareAcronym{MSE}{
	short=MSE,
	long=mean squared error
}
\DeclareAcronym{NMSE}{
	short=NMSE,
	long=normalized mean squared error
}
\DeclareAcronym{NLOS}{
	short=NLOS,
	long=non-line-of-sight
}

\DeclareAcronym{NLP}{
	short=NLP,
	long=natural language processing
}

\DeclareAcronym{NR}{
	short=NR,
	long=new radio
}
\DeclareAcronym{OFDM}{
	short=OFDM,
	long=orthogonal frequency-division multiplexing
}
\DeclareAcronym{OMP}{
	short=OMP,
	long=orthogonal matching pursuit
}
\DeclareAcronym{PDP}{
	short=PDP,
	long=power delay profiles
}
\DeclareAcronym{PO}{
	short=PO,
	long=phase offset
}
\DeclareAcronym{ppm}{
	short=ppm,
	long=parts-per-million
}
\DeclareAcronym{RF}{
	short=RF,
	long=radio frequency
 }
\DeclareAcronym{RMS}{
	short=RMS,
	long=root-mean-square
}
\DeclareAcronym{RPE}{
	short=RPE,
	long=relative precoding efficiency
}
\DeclareAcronym{RSU}{
	short=RSU,
	long=roadside unit
}
\DeclareAcronym{RTT}{
	short=RTT,
	long=round trip time
}
\DeclareAcronym{RX}{
	short=RX,
	long=receiver
}

\DeclareAcronym{RSRP}{
	short=RSRP,
	long=reference signal received power
 }

\DeclareAcronym{SNR}{
	short=SNR,
	long=signal-to-noise ratio
}
\DeclareAcronym{SVD}{
	short=SVD,
	long=singular value decomposition
}
\DeclareAcronym{SLAM}{
	short=SLAM,
	long=simultaneous localization and mapping
}
\DeclareAcronym{SBL}{
	short=SBL,
	long=sparse Bayesian learning
}
\DeclareAcronym{SOMP}{
	short=SOMP,
	long=simultaneous orthogonal matching pursuit 
}

\DeclareAcronym{TCN}{
	short=TCN,
	long=temporal convolutional network
}
\DeclareAcronym{ToF}{
	short=ToF,
	long=time of flight
}
\DeclareAcronym{TX}{
	short=TX,
	long=transmitter
}

\DeclareAcronym{TDoA}{
	short=TDoA,
	long=time-difference-of-arrival
}

\DeclareAcronym{ToA}{
	short=ToA,
	long=time-of-arrival
}
\DeclareAcronym{TO}{
	short=TO,
	long=timing offset
}

\DeclareAcronym{UL}{
	short=UL,
	long=uplink
}
\DeclareAcronym{ULA}{
	short=ULA,
	long=uniform linear array 
}
\DeclareAcronym{URA}{
	short=URA,
	long=uniform rectangular array 
}
\DeclareAcronym{V2I}{
	short=V2I,
	long=vehicle-to-infrastructure
}
\DeclareAcronym{V2V}{
	short=V2V,
	long=vehicle-to-vehicle
}
\DeclareAcronym{V2X}{
	short=V2X,
	long=vehicle-to-everything
}
\DeclareAcronym{VRU}{
	short=VRU,
	long=vulnerable road user
}
\DeclareAcronym{WLS}{
	short=WLS,
	long=weighted least squares
}
\DeclareAcronym{ViT}{
	short=ViT,
	long=vision transformer
}
\DeclareAcronym{MIM}{
	short=MIM,
	long=masked image modeling
}
\DeclareAcronym{MPC}{
	short=MPCs,
	long=multipath components
}
\DeclareAcronym{RSS}{
	short=RSS,
	long=received signal strength
}
\DeclareAcronym{PINN}{
	short=PINN,
	long=physics-informed neural network
}
\DeclareAcronym{CKM}{
	short=CKM,
	long=channel knowledge map
}